\documentclass{aa}
\usepackage{txfonts}
\usepackage{natbib}
\usepackage{amssymb}
\usepackage{amsmath}
\bibpunct{(}{)}{;}{a}{}{,}
\usepackage{epsf}
\usepackage{graphicx}
\usepackage{epsfig}
\usepackage{graphics}
\usepackage{subfigure}
\usepackage[english]{babel}
\usepackage{rotating}
\usepackage{color}

\newcommand{\teff}{$T_{\rm{eff}}$}
\newcommand{\logg}{$\log g$}
\newcommand{\lL}{\ifmmode \log \frac{L}{L_{\sun}} \else $\log \frac{L}{L_{\sun}}$\fi}

\newcommand{\vsini}{$V$~sin$i$}
\newcommand{\vinf}{$v_{\infty}$}

\newcommand{\vmac}{$v_{\rm mac}$}
\newcommand{\vesc}{$v_{\rm esc}$}
\newcommand{\kms}{km~s$^{-1}$}
\newcommand{\msun}{M$_{\sun}$}
\newcommand{\rsun}{R$_{\sun}$}

\begin{document}

\title{Properties of six short-period massive binaries: a study of the effects of binarity on surface chemical abundances\thanks{Based on observations made with the SOPHIE spectrograph on the 1.93-m telescope at Observatoire de Haute-Provence (OHP, CNRS/AMU), France}}
\author{F. Martins\inst{1}
\and L. Mahy\inst{2,3}\fnmsep\thanks{F.R.S.-FNRS Postdoctoral researcher}
\and A. Herv\'e\inst{1}
}
\institute{LUPM, Universit\'e de Montpellier, CNRS, Place Eug\`ene Bataillon, F-34095 Montpellier, France  \\
\and
Space sciences, Technologies, and Astrophysics Research (STAR) Institute, Universit\'e de Li\`ege, Quartier Agora, Bât. B5c, All\'ee du 6 aout, 19c, 4000, Li\`ege, Belgium\\
\and
Instituut voor Sterrenkunde, KU Leuven, Celestijnenlaan 200D, Bus 2401, B-3001 Leuven, Belgium
}

\offprints{Fabrice Martins\\ \email{fabrice.martins@umontpellier.fr}}

\date{Received / Accepted }

\abstract
{A significant fraction of massive stars are found in multiple systems. The effect of binarity on stellar evolution is poorly constrained. In particular, the role of tides and mass transfer on surface chemical abundances is not constrained observationally. }
{The aim of this study is to investigate the effect of binarity on the stellar properties and surface abundances of massive binaries.}
{We perform a spectroscopic analysis of six Galactic massive binaries. The spectra of individual components are obtained from a spectral disentangling method and are subsequently analyzed by means of atmosphere models. The stellar parameters and CNO surface abundances are determined. }
{Most systems are made of main-sequence stars. Three systems are detached, two are in contact and no information is available for the sixth system. For eleven out of the twelve stars studied the surface abundances are only mildly affected by stellar evolution and mixing. They are not different from those of single stars, within the uncertainties. The secondary of XZ~Cep is strongly chemically enriched. Considering previous determinations of surface abundance in massive binary systems suggests that the effect of tides on chemical mixing is limited, whereas mass transfer and removal of outer layers of the mass donor leads to the appearance of chemically processed material at the surface, although this is not systematic. The evolutionary masses of the components of our six systems are on average 16.5\% higher than the dynamical masses. Some systems seem to have reached synchronization, while others may still be in a transitory phase.}
{}

\keywords{Stars: early-type -- Stars: atmospheres -- Stars: fundamental parameters -- Stars: abundances -- Stars: binaries: general}

\authorrunning{Martins et al.}
\titlerunning{Properties of massive binaries}

\maketitle

\section{Introduction}
\label{s_intro}

The evolution of massive stars depends on several physical processes, the main ones being mass loss and rotation. Mass loss removes material from the outer layers, reducing the mass of the star and thus affecting its internal structure and luminosity \citep{cm86}. Main-sequence massive stars rotate on average faster than any other main-sequence star. Their average projected rotational velocity is found to be between 100 and 150 \kms\ depending on metallicity \citep{pg09,hg10,ra13}. Rotation flattens the star, triggering the transport of angular momentum and chemical species from the core to the surface \citep{mm00}. Rotation also modifies mass loss rates.
Other processes affect the evolution of massive stars. The presence of a strong dipolar magnetic field \citep[present in about 7\% of OB stars,][]{gru17} can alter both mass loss and rotation \citep{ud09,meynet11}. The presence of a companion can also modify the evolution of a star through tides and mass transfer \citep[e.g.][]{mandel16}.

A number of massive OB stars are found in binary systems. However the fraction of multiple systems is not fully established and varies depending on samples and environment. \citet{kob14} reported that 35\% (55\% after corrections for biases) of the 128 OB stars they studied in Cyg~OB2 definitely have a companion ; \citet{sana12} found a binary fraction of 56\% (69\% after corrections) for six Galactic open clusters and a total of 71 systems (40 definite binaries); \citet{sana14} detected a companion in 53\% of a sample of 96 Galactic O stars and claimed a binary fraction of 91\% accounting for known systems not detected in their survey (many systems having large separations though); in the Large Magellanic Cloud, \citet{sana13} determined an observed fraction of 35\% (corrected to 51\%) among 360 O stars (126 definite binaries); a minimum fraction of 21\% was determined by \citet{mahy13} for four Cygnus OB associations (19 systems, 4 definite binaries). In addition, \citet{kob14} and \citet{sana12} found a rather uniform distribution of mass ratios and a weak decrease of the period distribution among definite binaries. The latter distribution extends up to several thousands of days.

The presence of a companion around a significant number of OB stars raises the question of its effects on stellar evolution. Interactions are expected in close systems. For wide separations, both components evolve as single stars. There are two main categories of effects caused by the presence of a close companion: those due to tides and those due to mass transfer. The former trigger energy exchange between the system and its components. The consequence is a change of the internal structure under the influence of energy dissipation (and thus heating) through radiative damping, viscous friction and (gravito-)inertial waves \citep{zahn89a,mathisremus13}. The geometry of the star and its internal rotation profile can also be modified. Hence the transport of angular momentum and chemical species are affected. In practice, these effects lead to synchronization of the rotational and orbital periods \citep{zahn77}. This can spin-up or spin-down individual components and thus amplify or reduce the effects of rotation on their evolution. 
Mass transfer also drastically affects the evolution of massive stars. Due to the mass-luminosity relation, a decrease of mass implies a reduction of luminosity, and thus a change of the path in the Hertzsprung-Russell diagram \citep{wellstein01}. The mass donor loses its external layers, revealing at the surface hotter internal regions with different chemical properties. The mass gainer is polluted by material from the donor and its envelope is mixed with new material \citep{langer08}. Together with mass, angular momentum is exchanged. This affects the rotational properties of both stars, in a way that depends on the efficiency of mass transfer. The outcome of mass transfer depends on the orbital properties of the system (mass ratio, separation) and on the intrinsic evolution of the components.

Stellar evolution in (massive) binaries is thus complex. Constraints from analysis of dedicated systems are therefore highly requested. One of the expectations of binary evolution is that surface abundances are modified compared to single stars. Indeed, mixing processes due to rotation depend on the star's internal structure which is affected by tides \citep{demink09}. Contamination by accretion in mass transfer episodes obviously affects surface chemical patterns \citep{langer08}. In this case, the mass donor loses its external layers so that its surface is physically moved to deeper layers where chemical composition corresponds to more processed material. These effects of binarity on surface abundances are one of the possibilities quoted to explain peculiar abundance patterns. For instance, the nitrogen-rich stars with low projected rotational velocity (as well as their counterparts, nitrogen-poor stars with high \vsini) reported by \citet{hunter08} -- see also \citet{grin17} -- may be the result of such interactions according to the authors. However the difference between the chemical patterns of binary and single stars has not been established observationally. The attempt performed by \citet{garland17} on a sample of B-type binaries did not show any clear difference, but only the primary component of their systems was analyzed.

Here we present a study of six Galactic massive binaries. Our goal is to determine the stellar parameters and surface CNO abundances of their components in order to investigate whether they are different from presumably single stars. Section \ref{s_obs} describes our sample and the spectroscopic observations. Data analysis and the associated results are gathered in Sect.\ \ref{s_mod}. Results are discussed in Sect.\ \ref{s_disc}. Finally we summarize our conclusions in Sect.\ \ref{s_conc}.

\section{Sample and observations}
\label{s_obs}

In order to test the effects of binarity on the properties and evolution of massive stars, we selected six short period systems ($P < 6$ days) from the compilation of \citet{gies03}. All systems except DH~Cep are eclipsing binaries. The targets presented in this study were observed at Observatoire de Haute-Provence with the \emph{SOPHIE} instrument \citep{bouchy13} during the nights of 22$^{nd}$ to 26$^{th}$ August 2013. \emph{SOPHIE} delivers high spectral resolution \'echelle spectra in the 3900-6900\AA\ wavelength range. The main properties of the observed binaries are listed in Table~\ref{tab_obs}. The High Efficiency mode of \emph{SOPHIE}, corresponding to a spectral resolution $R = 39000$, was used. The spectra were obtained in 15 to 40 minutes depending on the target's brightness and have a signal-to-noise ratio close to 200 in the continuum. The observing strategy (number of observations per system) was adjusted each night on order to ensure an optimal coverage of the phases of maximum separation between the systems' components as requested for a good spectral disentangling. Data reduction was performed automatically by the \emph{SOPHIE} pipeline \citep{bouchy09} adapted from the ESO/HARPS software.

\begin{table}
\begin{center}
\caption{Observational information: Target name, spectral type, period, V magnitude, and number of spectra (Nb) obtained for each system.} \label{tab_obs}
\begin{tabular}{lccccc}
\hline
Star        & ST                  & Period  &  V    & Nb \\    
            &                     & [d]     &       &    \\
\hline
DH Cep	    & O5.5V-III+O6V-III	  & 2.11095 &  8.61 & 11 \\
V382 Cyg    & O6.5V((f))+O6V((f)) & 1.88555 &  8.65 & 10 \\
Y~Cyg       & O9V+O9.5V           & 2.99633 &  7.32 & 9 \\
V478~Cyg    & O9.5V+O9.5V         & 2.88086 &  8.68 & 10 \\
XZ Cep	    & O9.5V+B1III	  & 5.09725 &  8.51 & 6 \\
AH Cep	    & B0.2V+B2V		  & 1.77473 &  6.88 & 15 \\
\hline
\end{tabular}
\end{center}
\end{table}

\section{Modelling and spectroscopic analysis}
\label{s_mod}

\subsection{Orbital solutions and spectral disentangling}
\label{orbit}

\begin{figure*}
     \centering
     \subfigure{
          \includegraphics[width=.45\textwidth]{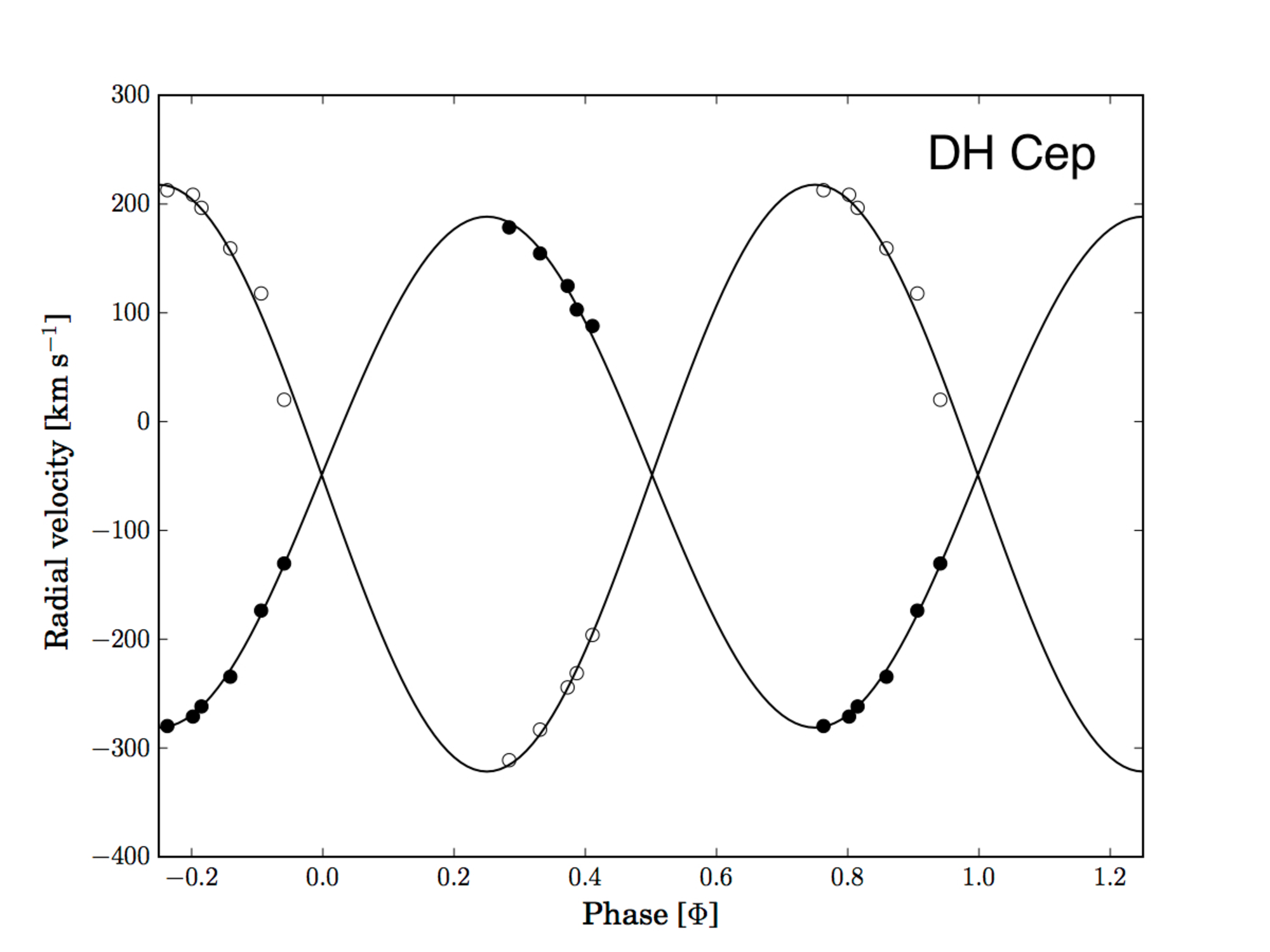}}
     \hspace{0.2cm}
     \subfigure{
          \includegraphics[width=.45\textwidth]{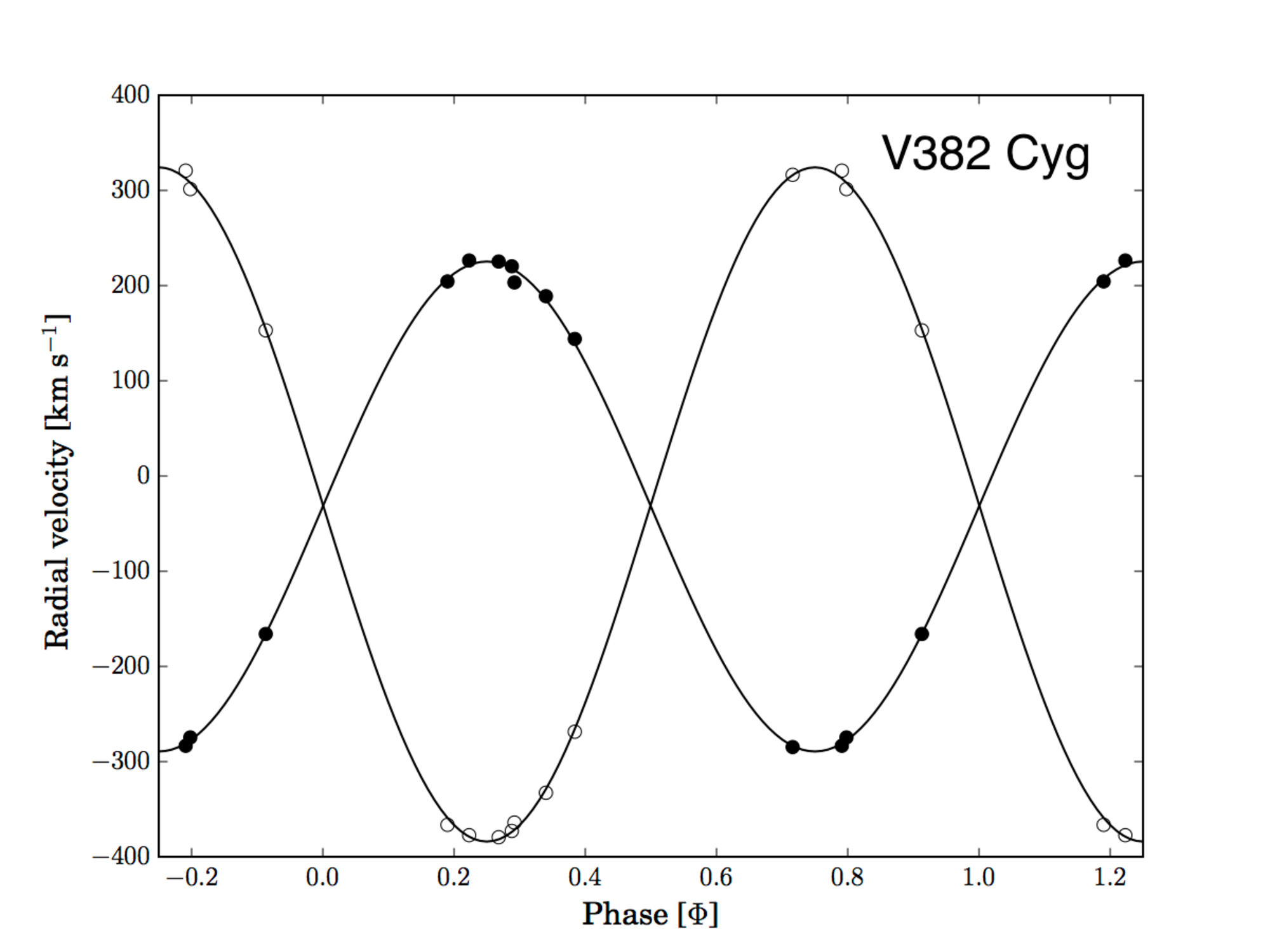}}\\
     \subfigure{
          \includegraphics[width=.45\textwidth]{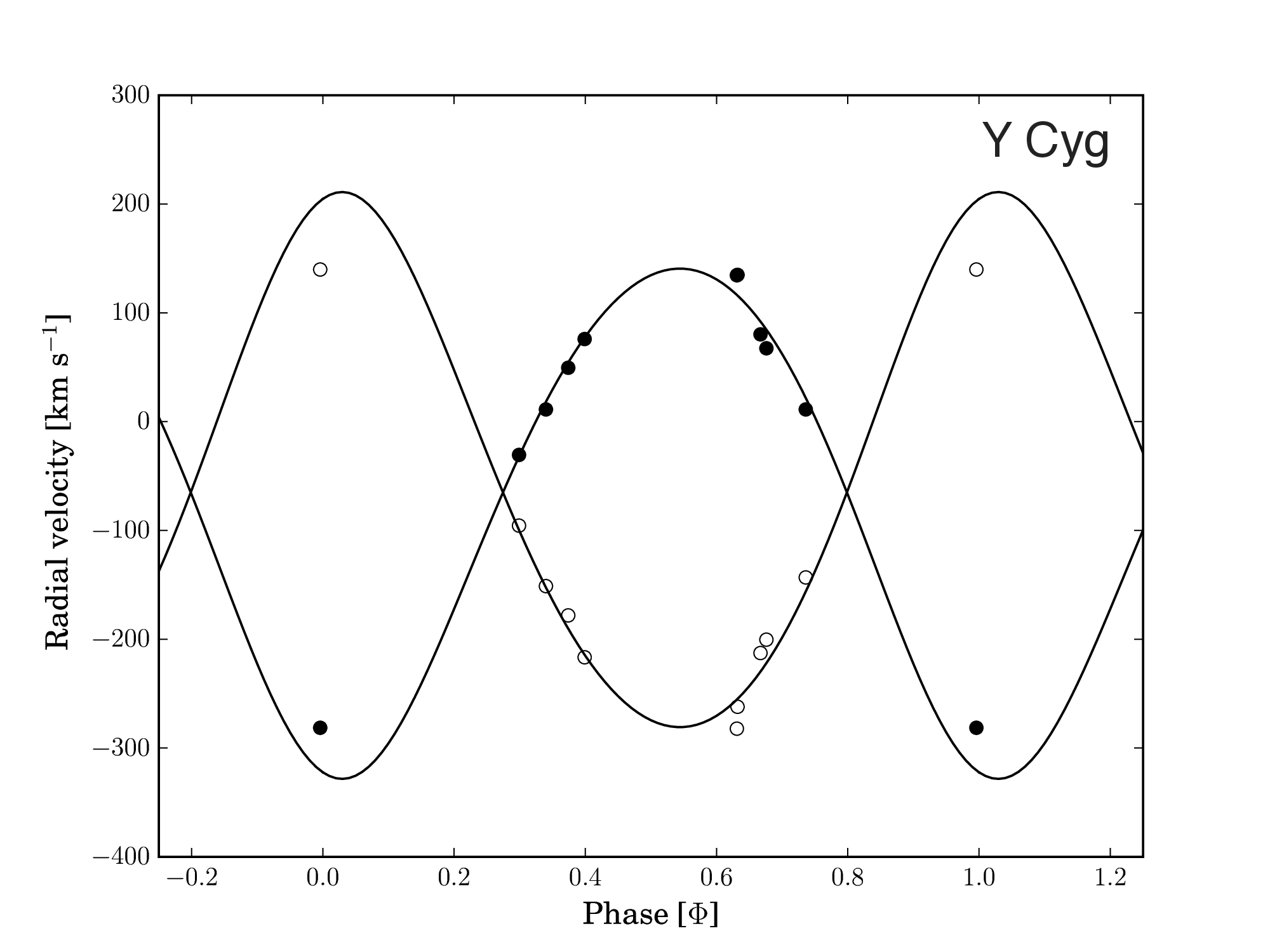}}
     \hspace{0.2cm}
     \subfigure{
          \includegraphics[width=.45\textwidth]{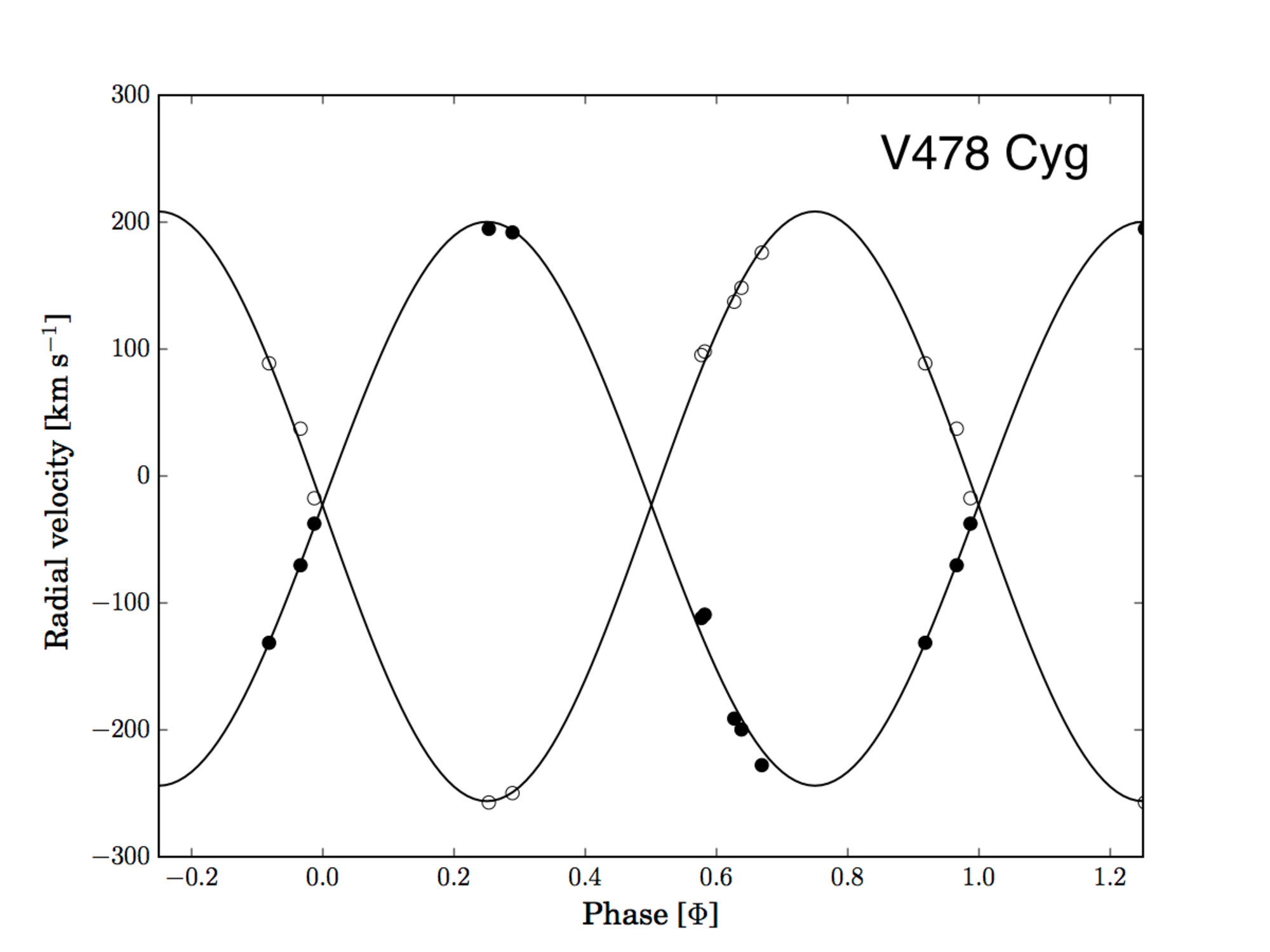}}\\
     \subfigure{
          \includegraphics[width=.45\textwidth]{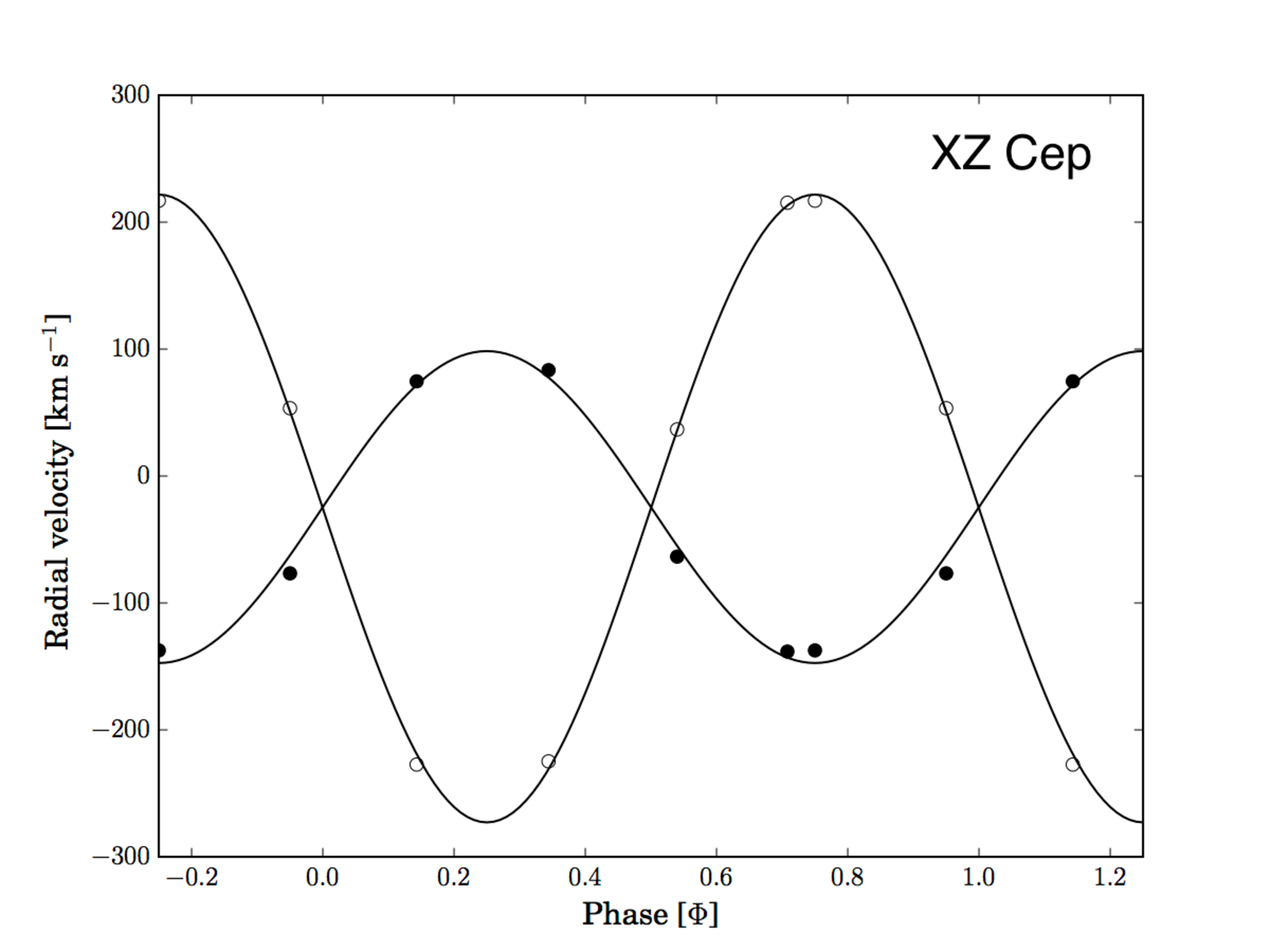}}
     \hspace{0.2cm}
     \subfigure{
          \includegraphics[width=.45\textwidth]{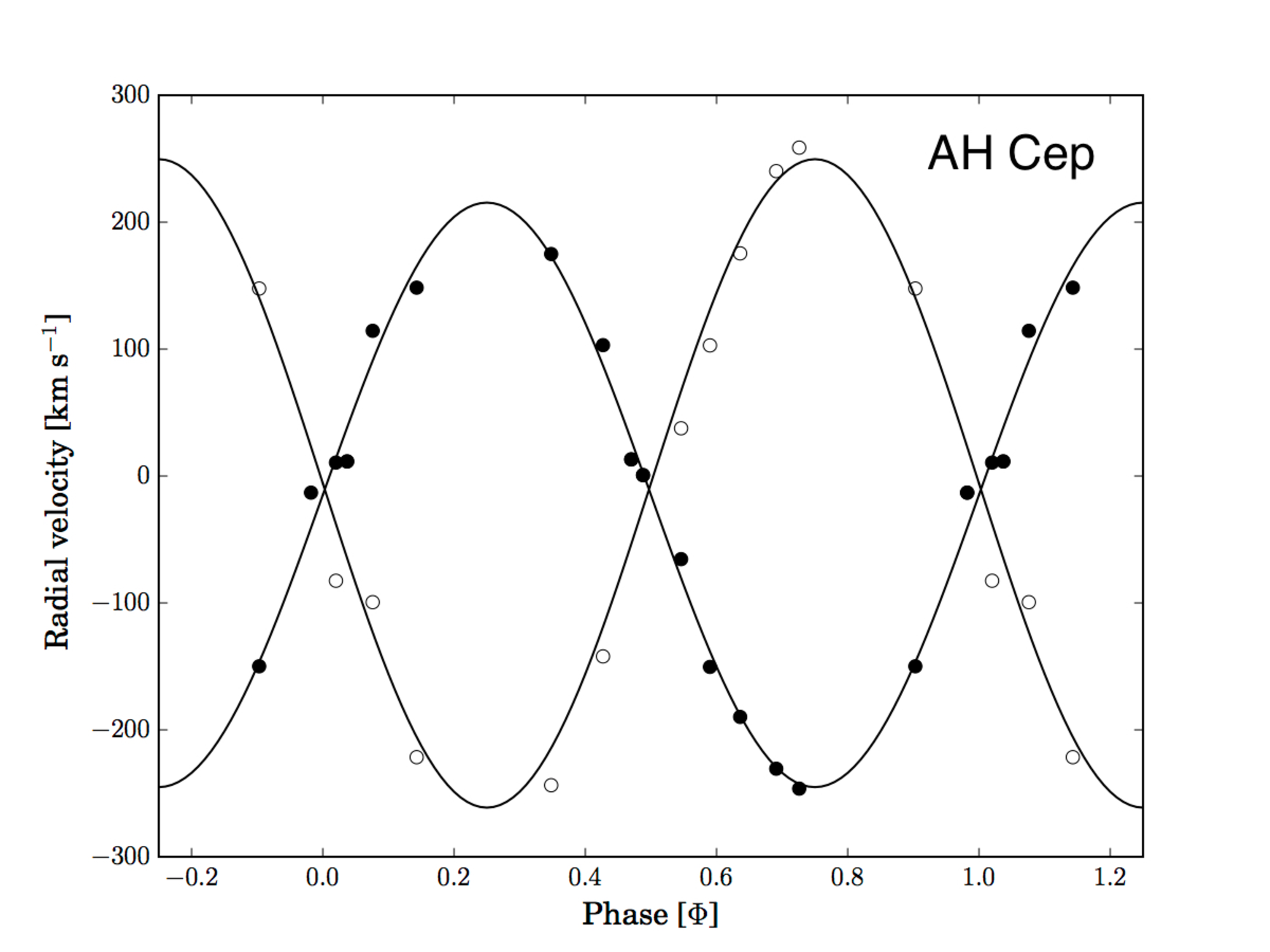}}
     \caption{Orbital solutions (solid) lines and measured radial velocities (circles) for the primary (filled symbols) and secondary (open circle) of each system.}
     \label{fig:orbital}
\end{figure*}

To compute the orbital solution of each system, we first measured the radial velocities of each component. For this purpose, we fitted the spectral lines of helium with the highest ionization stage. The higher the ionization stage, the closer to the photosphere the lines are formed. The computations of the radial velocities (RVs) were done by adopting the rest wavelength from \citet{conti77}. We then used the Li\`ege Orbital Solution Package (LOSP\footnote{LOSP is developed and maintained by H. Sana. The algorithm is based on the generalization of the SB1 method of \citet{wolfe67} to the SB2 case along the lines described in \citet{rauw00} and \citet{sana06}. It is available at \url{http://www.stsci.edu/~hsana/losp.html}.}) to determine the SB2 orbital solution of the system. Since the orbital periods of these systems are well known, we decided to fix them in the computations. Except for Y\,Cyg, the other systems have all a circular orbit. The orbital solutions are displayed in Fig.\,\ref{fig:orbital}. The corresponding parameters are gathered in Table\,\ref{tab_orb}.

We then used the orbital parameters as input of our spectral disentangling code. We applied the Fourier approach of \citet{Hadrava95} to separate the spectral contributions of each component in the different systems. This method uses the Nelder \& Mead’s Downhill Simplex method on the multidimensional parameter space to reach the best $\chi^2$ fit between the recombined component spectra and the observed data. Discussions about the spectral disentangling method used in this paper can be found in \citet{pavlovski10} and in \citet{mahy17b}. Because the eclipses are not total in the different systems, the output composite spectra must be renormalized and corrected for the brightness ratios. A first approximation of the brightness ratio was estimated by computing the equivalent widths of several spectral lines (\ion{Si}{iv}~4089, \ion{He}{i}~4143, \ion{He}{i}~4471, \ion{He}{ii}~4542, \ion{O}{iii}~5592) and by comparing these values to those calculated from synthetic spectra corresponding to stars with the same spectral types as the components of our sample. Then, the brightness ratios have been iteratively adjusted by computing them from the ratios between the luminosities of the two components (luminosities that have been obtained from the effective temperature and the radius of each star).

\begin{table*}
\begin{center}
\caption{Orbital solution for the sample stars. 1-$\sigma$ errors are given.} 
\label{tab_orb}
\begin{tabular}{lrrrrrr}
\hline\hline
& \multicolumn{2}{c}{AH\,Cep} & \multicolumn{2}{c}{DH\,Cep} & \multicolumn{2}{c}{XZ\,Cep}\\
& Primary        & Secondary  & Primary        & Secondary  & Primary        & Secondary \\
\hline
$P$~[day] & \multicolumn{2}{c}{1.774727 (fixed)} & \multicolumn{2}{c}{2.11095 (fixed)} & \multicolumn{2}{c}{5.097253 (fixed)}\\
$e$       &  \multicolumn{2}{c}{$0.0$ (fixed)}   &  \multicolumn{2}{c}{$0.0$ (fixed)}       &  \multicolumn{2}{c}{$0.0$ (fixed)}\\
$\omega$~[$\degr$] & \multicolumn{2}{c}{--}      &  \multicolumn{2}{c}{--} &  \multicolumn{2}{c}{--}\\
$T_0$~[HJD~$-$~2\,450\,000]  & \multicolumn{2}{c}{6526.783 $\pm$ 0.008}  & \multicolumn{2}{c}{6525.564 $\pm$ 0.006}  & \multicolumn{2}{c}{6522.656 $\pm$ 0.014}\\
$q~(M_1/M_2)$ & \multicolumn{2}{c}{1.132 $\pm$ 0.056} & \multicolumn{2}{c}{1.149 $\pm$ 0.020} & \multicolumn{2}{c}{2.013 $\pm$ 0.092}\\
$\gamma$~[\kms]           & $-14.69 \pm 5.29$   & $-13.43 \pm 5.53$   & $-46.43 \pm 2.47$   & $-51.95 \pm 2.65$    & $-24.47 \pm 3.49$  & $-25.58 \pm 4.39$ \\
$K$~[\kms]                &  236.46 $\pm$ 7.33  & 267.57 $\pm$ 8.29   & 234.81 $\pm$ 3.97   &  269.70 $\pm$ 4.56   &  122.86 $\pm$ 3.49 & 247.29 $\pm$ 7.03 \\
$a \sin i$~[$R_{\odot}$]   & 8.29 $\pm$ 0.26      & 9.38 $\pm$ 0.29    & 9.79 $\pm$ 0.17     & 11.24 $\pm$ 0.19     & 12.37 $\pm$ 0.35    & 24.89 $\pm$ 0.71 \\
$M \sin^3 i$~[\msun]      & 12.49 $\pm$ 0.78     & 11.04 $\pm$ 0.66   & 15.01 $\pm$ 0.66    & 13.07 $\pm$ 0.57     & 17.88 $\pm$ 1.14   & 8.88 $\pm$ 0.46 \\
rms~[\kms] & \multicolumn{2}{c}{$14.36$}  & \multicolumn{2}{c}{$8.78$}  & \multicolumn{2}{c}{$9.09$}\\
\hline
 & \\
& \multicolumn{2}{c}{V382\,Cyg} & \multicolumn{2}{c}{V478\,Cyg} & \multicolumn{2}{c}{Y\,Cyg}\\
& Primary        & Secondary  & Primary        & Secondary  & Primary        & Secondary \\
\hline
$P$~[day] & \multicolumn{2}{c}{1.885545 (fixed)} & \multicolumn{2}{c}{2.880867 (fixed)} & \multicolumn{2}{c}{2.9963316 (fixed)}\\
$e$       &  \multicolumn{2}{c}{$0.0$ (fixed)}   &  \multicolumn{2}{c}{$0.0$ (fixed)}       &  \multicolumn{2}{c}{$0.128 \pm 0.016$}\\
$\omega$~[$\degr$] & \multicolumn{2}{c}{--}      &  \multicolumn{2}{c}{--} &  \multicolumn{2}{c}{168.63 $\pm$ 4.95 }\\
$T_0$~[HJD~$-$~2\,450\,000]  & \multicolumn{2}{c}{6527.020 $\pm$ 0.003}  & \multicolumn{2}{c}{6525.763 $\pm$ 0.006}  & \multicolumn{2}{c}{6526.454 $\pm$ 0.041}\\
$q~(M_1/M_2)$ & \multicolumn{2}{c}{1.376 $\pm$ 0.009} & \multicolumn{2}{c}{1.046 $\pm$ 0.040} & \multicolumn{2}{c}{1.053 $\pm$ 0.017}\\
$\gamma$~[\kms] & $-32.04 \pm 1.30$ & $-29.88 \pm 1.60$ & $-21.87 \pm 3.80$ & $-23.87 \pm 3.86$  &  $-68.55 \pm 2.90$ & $-61.47 \pm 2.99$ \\
$K$~[\kms] &  257.32 $\pm$ 1.46 & 354.11 $\pm$ 2.01  &  222.11 $\pm$ 5.43 & 232.28 $\pm$ 5.67   & 232.41 $\pm$ 3.00  &  244.81 $\pm$ 3.16\\
$a \sin i$~[$R_{\odot}$] & 9.58 $\pm$ 0.05 & 13.19 $\pm$ 0.08 & 12.64 $\pm$ 0.31 & 13.22 $\pm$ 0.32  & 13.64 $\pm$ 0.18  & 14.37 $\pm$ 0.19\\
$M \sin^3 i$~[\msun] & 25.85 $\pm$ 0.38 & 18.78 $\pm$ 0.27  & 14.31 $\pm$ 0.72 & 13.68 $\pm$ 0.68   & 16.87 $\pm$ 0.53 & 16.02 $\pm$ 0.50\\
rms~[\kms] & \multicolumn{2}{c}{$5.55$}  & \multicolumn{2}{c}{$10.13$}  & \multicolumn{2}{c}{$12.41$}\\
\hline         
\end{tabular}
\tablefoot{For circular systems, $T_0$ refers to the time of the conjunction (primary in front) and for the eccentric system, $T_0$ refers to the periastron passage. $\gamma$, $K$ and $a \sin i$ denote the apparent systemic velocity, the semi-amplitude of the radial velocity curve and the projected separation between the center of the star and the center of mass of the binary system.}
\end{center}
\end{table*}

\subsection{Stellar parameters}
\label{s_param}

We used the Fourier transform method \citep{gray76,sergio07} to determine the projected rotational velocity \vsini. When the combined spectra showed sufficiently wide separations between both components of the system so that individual lines could be clearly isolated, we used directly the combined spectrum. Otherwise, we relied on the disentangled spectra. The uncertainty of our determinations is 10 \kms. We found that all stars have \vsini\ larger than 100 \kms. 

To estimate the amount of extra-broadening (macroturbulence) we used synthetic spectra (see below) with \teff\ and \logg\ consistent with the spectral type and the luminosity class of each star \citep{msh05,np14}. We convolved these spectra with a rotational profile adopting the \vsini\ derived from the Fourier transform method. We added an extra-broadening by means of a radial-tangential profile parameterized by a velocity \vmac. The comparison with the observed, disentangled spectrum of each component revealed that for most systems there was no need for an extra-broadening. We thus adopted \vmac\ = 0 \kms. Only for DH~Cep did we choose \vmac\ = 100 \kms\ in order to correctly reproduce the line wings.

\begin{figure}[]
\centering
\includegraphics[width=0.47\textwidth]{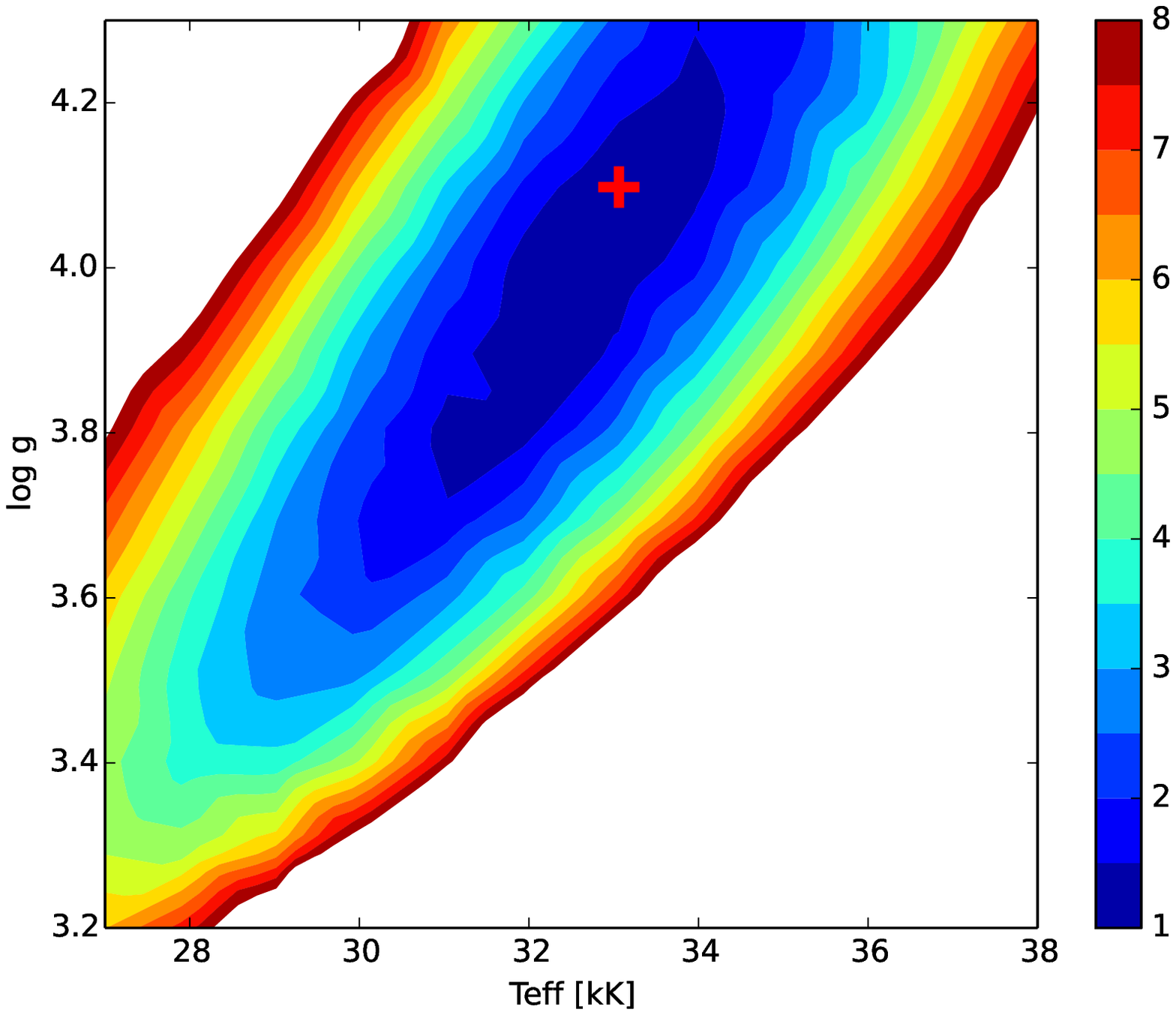}
\caption{Determination of \teff\ and \logg. Colors illustrate the value of $\chi^2$ for the fit of a synthetic spectrum with a given \teff--\logg\ combination to the observed disentangled spectrum of the primary component of Y~Cyg. The red cross indicates the position at which $\chi^2 = 1.0$. }
\label{fig_chi2}
\end{figure}

\teff\ and \logg\ were determined from the classical diagnostics: the helium ionization balance for \teff\ and the Balmer lines broadening for the surface gravity \citep[e.g.][]{liege11}. For B stars we replaced helium by silicon for the effective temperature determination. We relied on a grid of synthetic spectra computed with the code CMFGEN \citep{hm98}. Non-LTE atmosphere models were computed for stars with 21000 $<$ \teff\ $<$ 45000 K and 3.0 $<$ \logg\ $<$ 4.3. Luminosities were assigned according to the calibration of \citet{msh05} and mass loss rates were computed from \citet{vink01}. Wind terminal velocities were estimated from \vinf\ = 3.0 $\times$ \vesc\ \citep{garcia14}. A solar composition \citep{ga10} was assumed and the following elements were included: H, He, C, N, O, Ne, Mg, Si, S, Ar, Ca, Fe, Ni. Once atmosphere models were converged a formal solution of the radiative transfer equation lead to the synthetic emergent spectrum, which was computed assuming a microturbulent velocity ranging from 10 \kms\ at the photosphere to 0.1 $\times$ \vinf\ in the outermost parts of the atmosphere. For each star, a subset of the model grid (encompassing the values of \teff\ and \logg\ estimated from the spectral type) was convolved with the appropriate \vsini\ and \vmac\ as described above. The resulting synthetic spectra, correctly shifted in radial velocities, were subsequently compared to the disentangled spectrum of each target. The quality of the fit was quantified by means of a $\chi^2$ analysis renormalized to 1.0 for the best-fit combination of \teff\ and \logg. The output of such a process is illustrated in Fig.\ \ref{fig_chi2} in the case of the primary component of Y~Cyg. The uncertainties on \teff\ and \logg\ were estimated from such figures. In practice, \teff\ and \logg\ are correlated and giving a formal error on \teff\ and \logg\ is not correct. However, we decided to take the width of the $\chi^2 = 2.0$ contour as a representative error. These values are reported in Table \ref{tab_param}. 

Surface abundances of C, N and O were determined as in \citet{mimesO}: for a given set of \teff\ and \logg\ models with different C, N and O content were computed and subsequently compared to disentangled spectra. The goodness of fit was quantified by means of a $\chi^2$ analysis. In practice, we relied on the following lines:  

\begin{itemize}

\item for carbon: \ion{C}{iii}~4068-70, \ion{C}{iii}~4153, \ion{C}{iii}~4156, \ion{C}{iii}~4163, \ion{C}{iii}~4187, \ion{C}{ii}~4267,  \ion{C}{ii}~6578, and \ion{C}{ii}~6583.

\item for nitrogen: \ion{N}{ii}~3995, \ion{N}{ii}~4004, \ion{N}{ii}~4035, \ion{N}{ii}~4041, \ion{N}{iii}~4044, \ion{N}{ii}~4236, \ion{N}{ii}~4447, \ion{N}{iii}~4511, \ion{N}{iii}~4515, \ion{N}{iii}~4518, \ion{N}{iii}~4524, \ion{N}{ii}~4530, \ion{N}{ii}~4601, \ion{N}{ii}~4607, \ion{N}{ii}~4614,  \ion{N}{ii}~4621,  \ion{N}{ii}~4630, \ion{N}{ii}~4788, \ion{N}{ii}~4803, \ion{N}{ii}~5001, \ion{N}{ii}~5005  and \ion{N}{ii}~5680.

\item for oxygen: \ion{O}{ii}~3913, \ion{O}{ii}~3955, \ion{O}{ii}~4120, \ion{O}{ii}~4284, \ion{O}{ii}~4305, \ion{O}{ii}~4318, \ion{O}{ii}~4321, \ion{O}{ii}~4368, \ion{O}{ii}~4592, \ion{O}{ii}~4597, \ion{O}{ii}~4603, \ion{O}{ii}~4611, \ion{O}{ii}~4663, \ion{O}{ii}~4678, \ion{O}{ii}~4700, \ion{O}{ii}~4707, and \ion{O}{iii}~5592. 

\end{itemize}

\begin{figure*}[]
\centering
\includegraphics[width=0.42\textwidth]{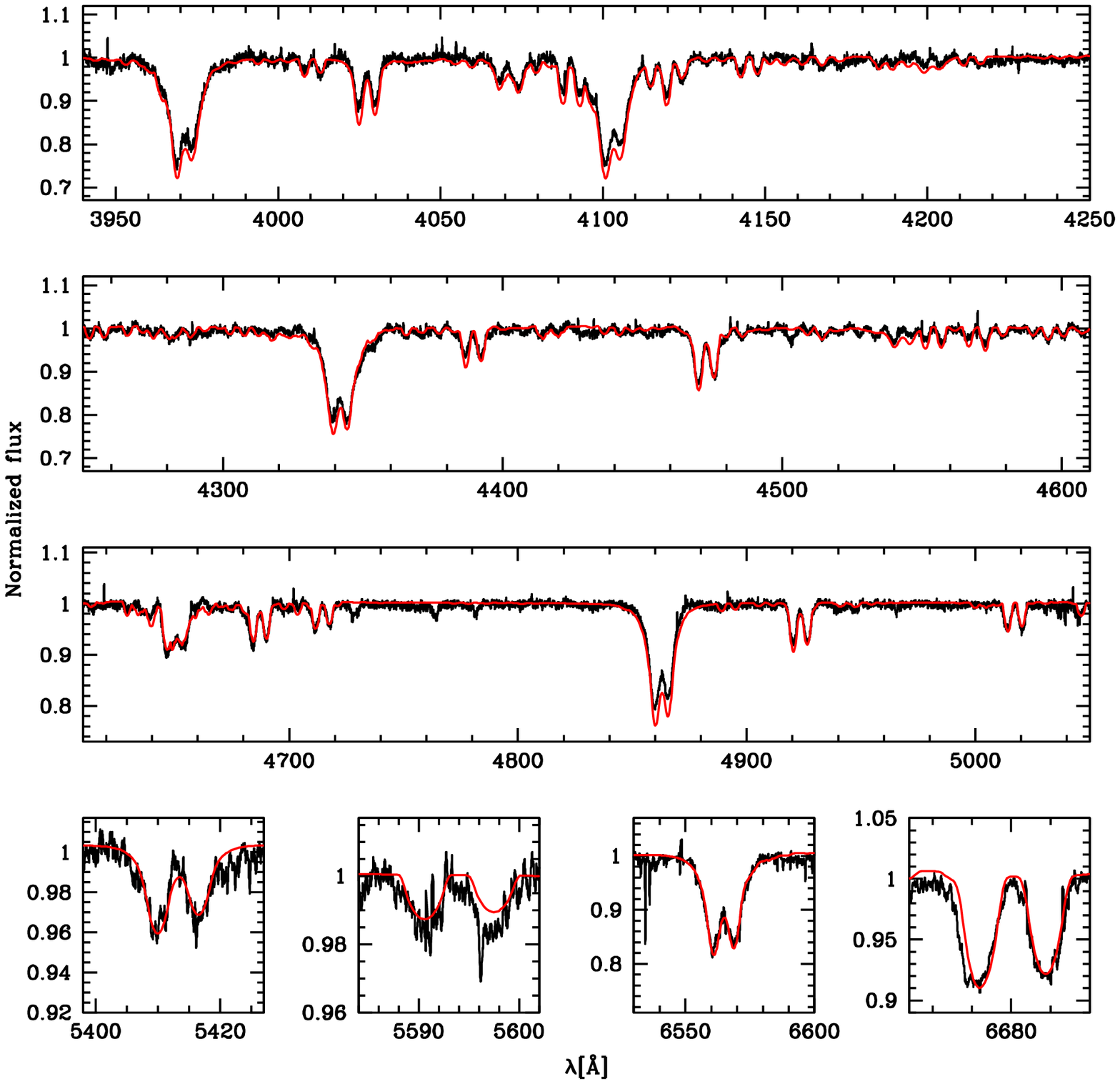}
\includegraphics[width=0.42\textwidth]{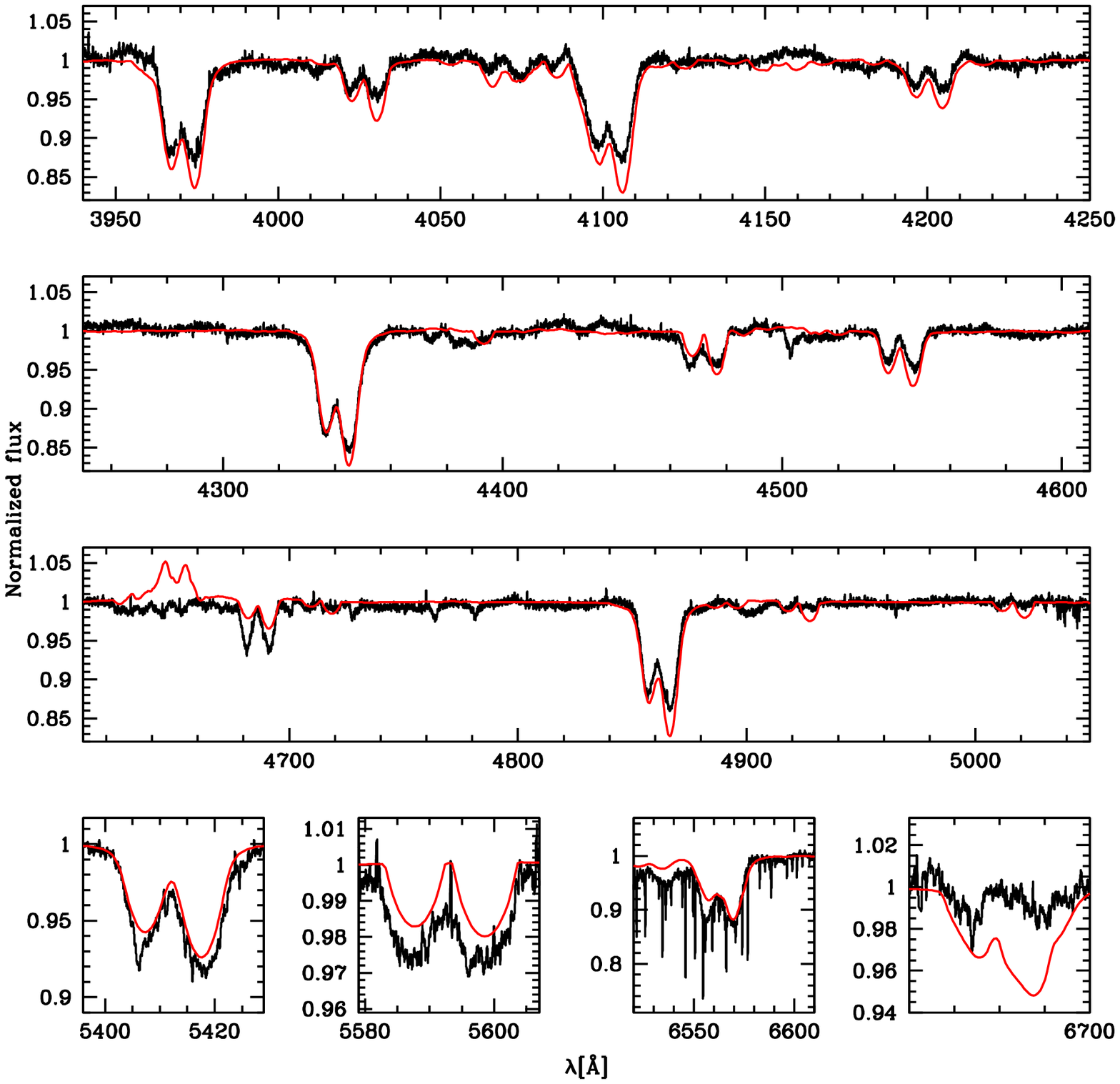}\\
\includegraphics[width=0.42\textwidth]{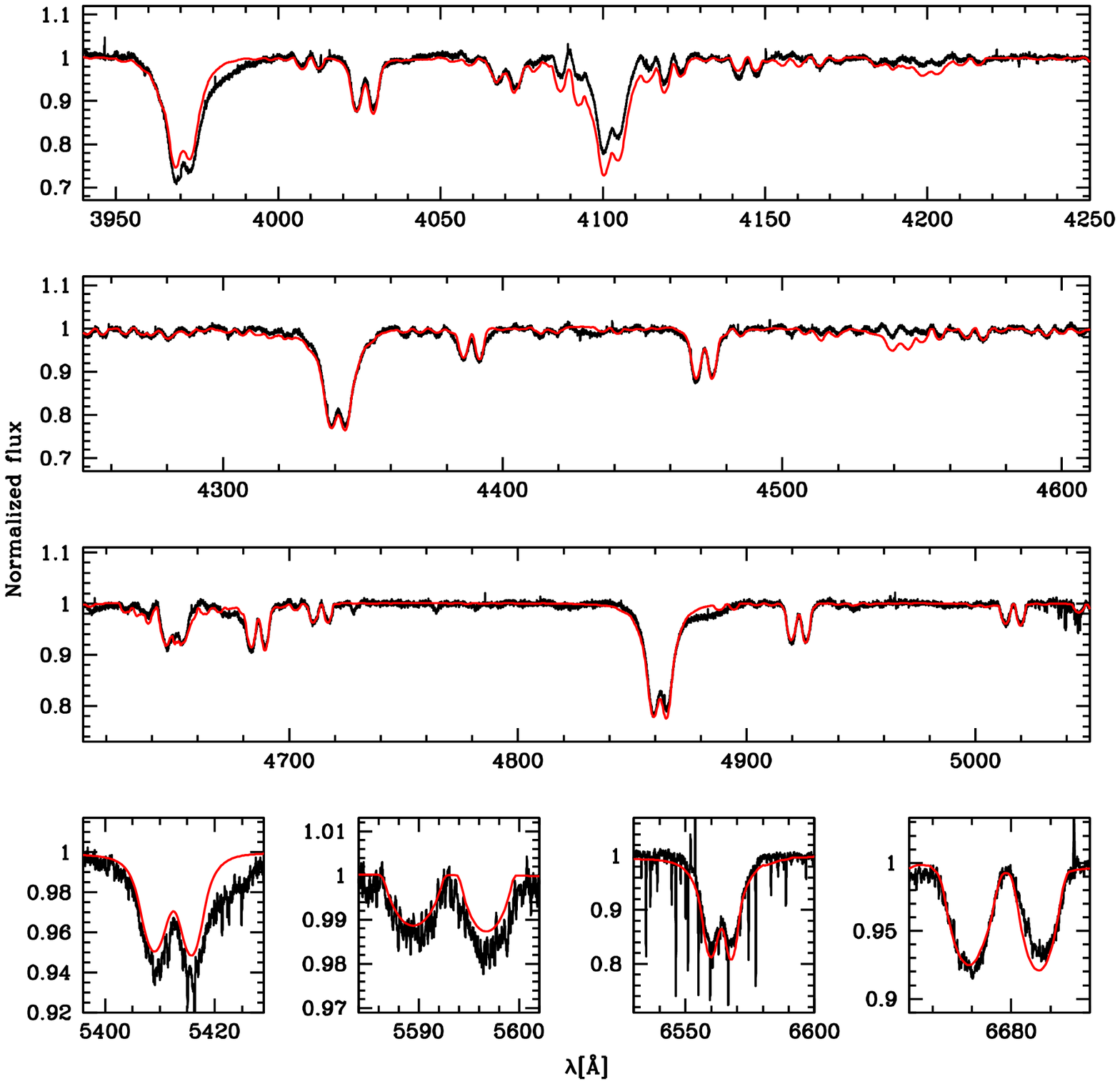}
\includegraphics[width=0.42\textwidth]{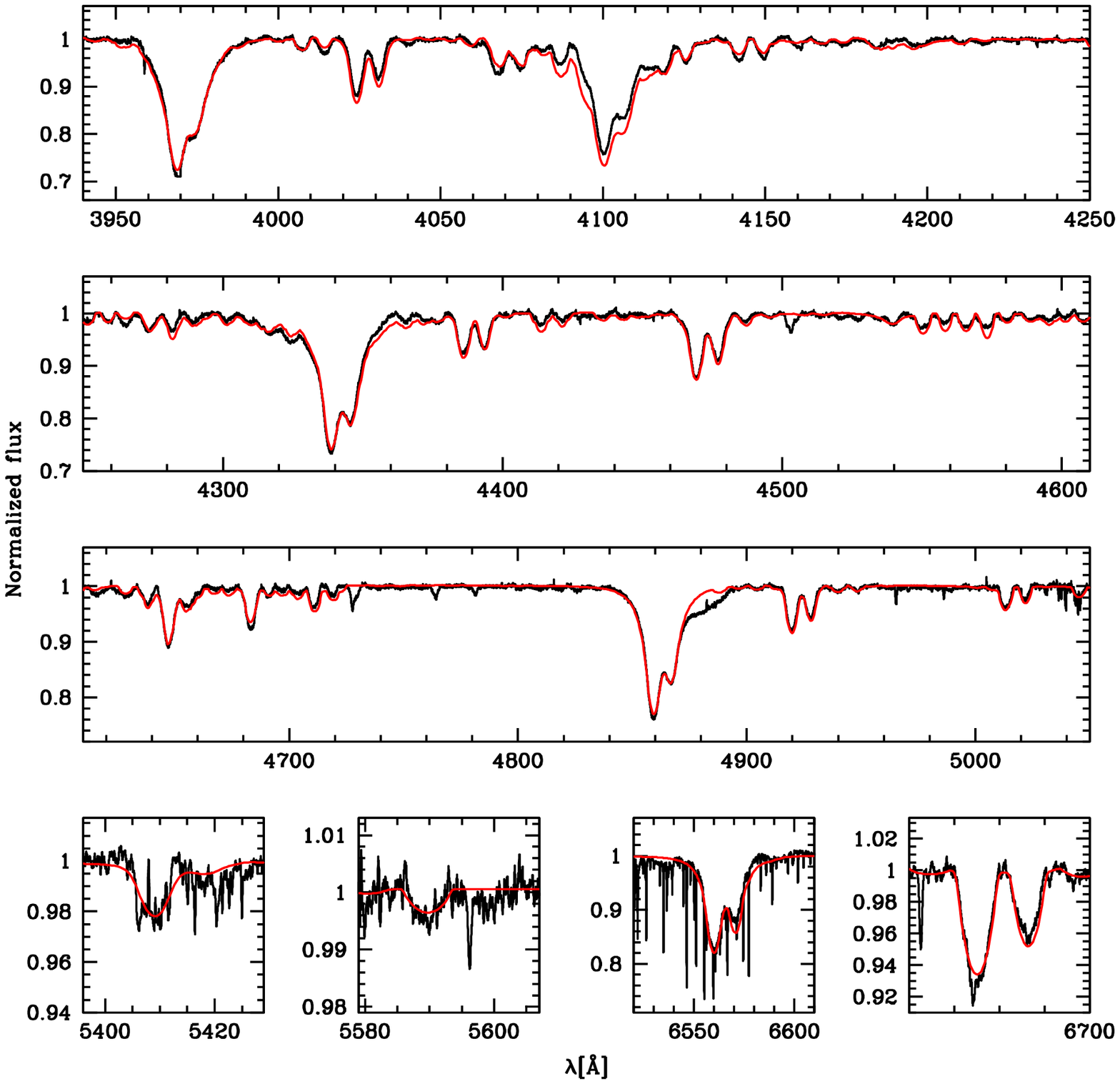}\\
\includegraphics[width=0.42\textwidth]{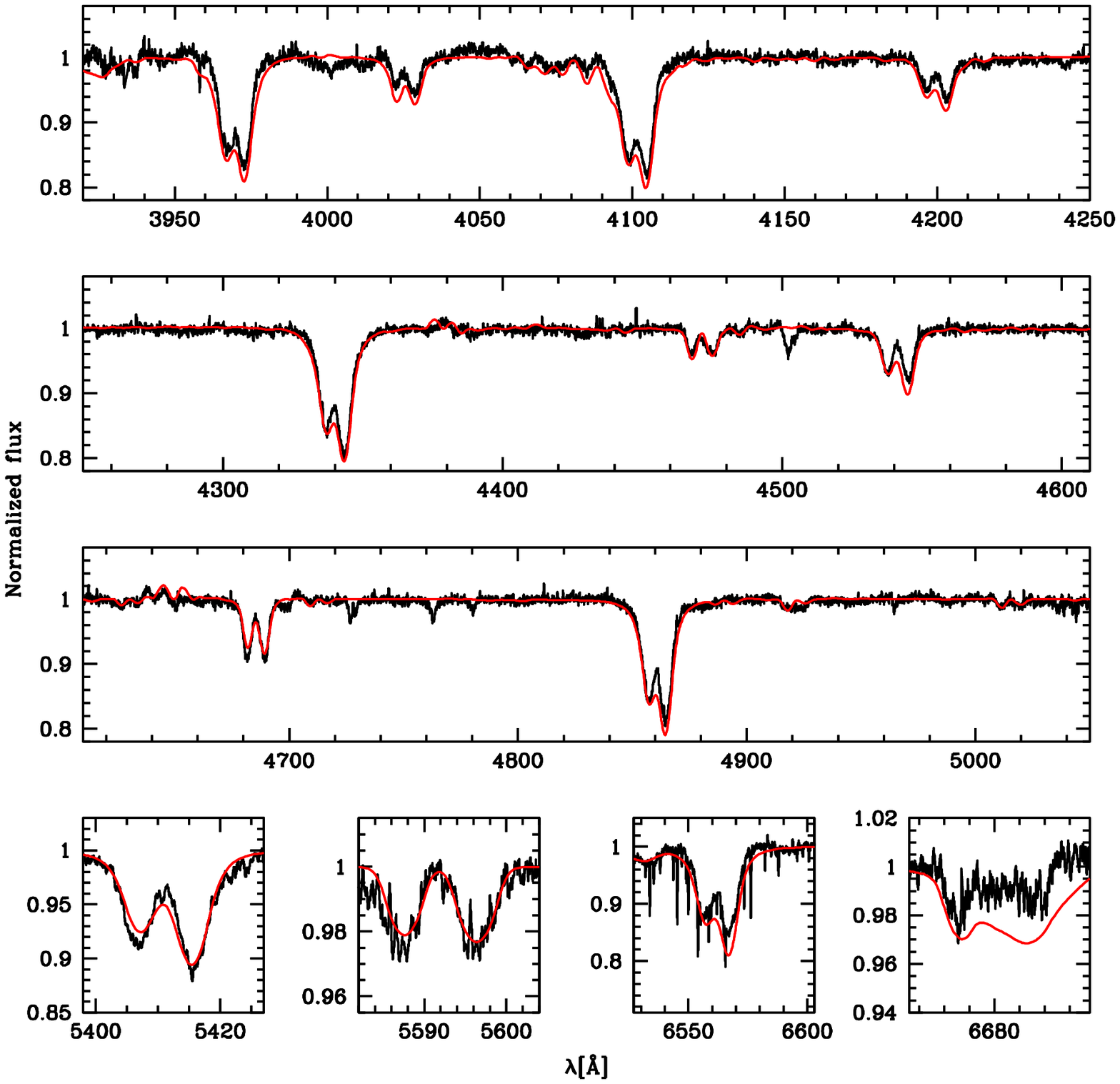}
\includegraphics[width=0.42\textwidth]{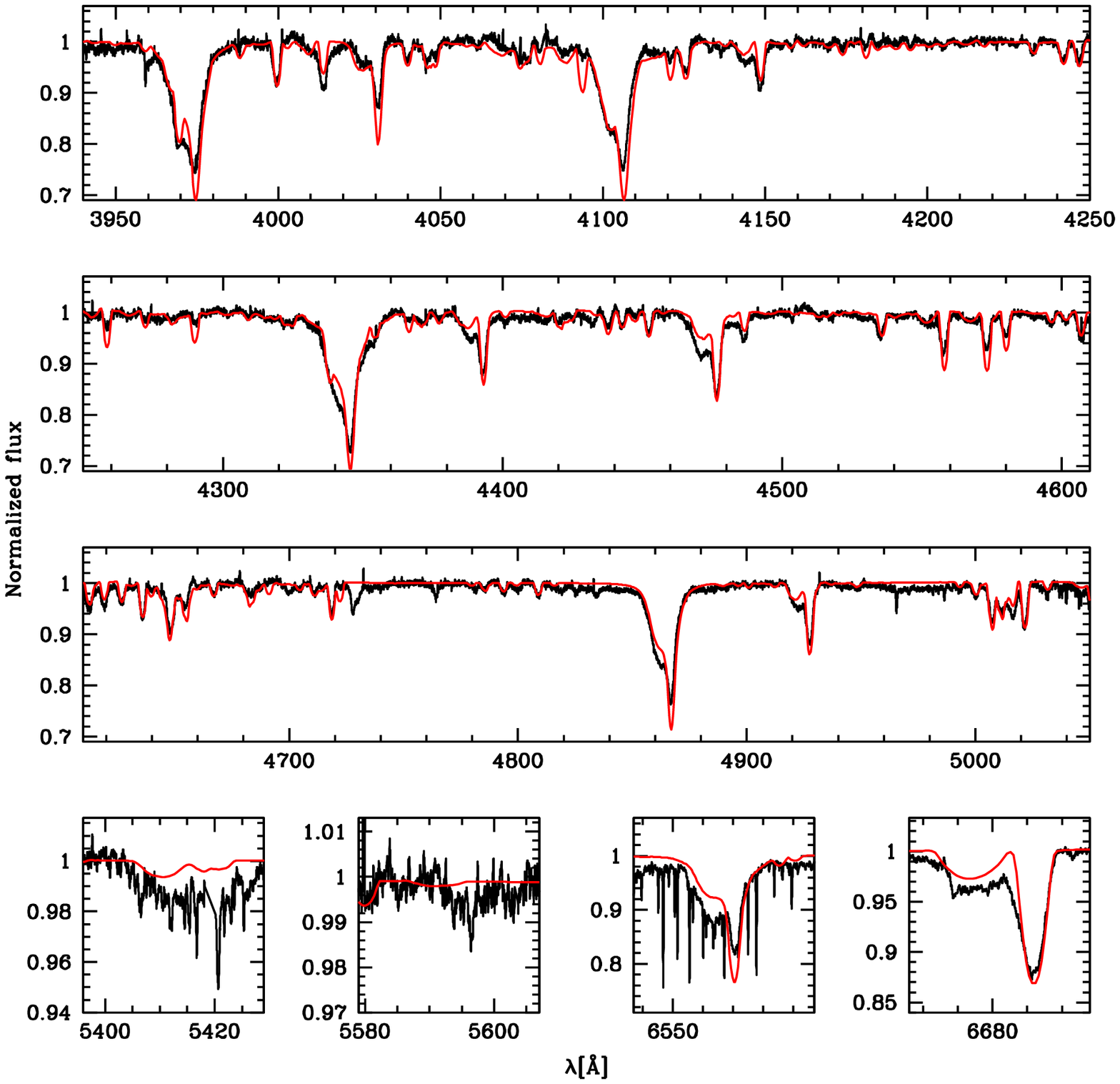}
\caption{Fit of one of the observed spectra of V478~Cyg (top left), V382~Cyg (top right), Y~Cyg (middle left), AH~Cep (middle right), DH~Cep (bottom left) and XZ~Cep (bottom right). The observed spectrum is shown in black and the best fit model in red. The latter takes into account the luminosity ratio and the radial velocities of both components.}
\label{fit_all}
\end{figure*}

\noindent The number of lines used for each star depends on the effective temperature and on the quality of the spectra. We adopted He/H=0.1 in our computations. No significant deviation from this value was observed in our fits (see Fig.\ \ref{fit_all}).

The derived parameters are given in Table~\ref{tab_param}. The final best-fit models are shown in Fig.\ \ref{fit_all}. For each system, the best fit of the individual components were co-added taking into account the luminosity ratio and the radial velocities determined at the date at which the comparison spectrum was obtained. In general, the quality of the fits is good. Notable problems are encountered around Balmer lines, especially H$\delta$, illustrating the difficulty to correctly disentangle broad lines that are never completely separated observationally. This is even worse when their wings contain lines from other elements.

\begin{table*}
\begin{center}
\caption{Stellar parameters} \label{tab_param}
\begin{tabular}{lcccccccccc}
\hline
Star         & \teff  &  \logg   &  \lL   &  \vsini   &  \vmac &  C/H        &  N/H &  O/H \\    
             & [kK]      &         &       &  [\kms]         & [\kms]       &  [10$^{-4}$] &  [10$^{-4}$] &  [10$^{-4}$] \\
\hline                       
DHCep-1      & 44$\pm$3.0   & 4.3$\pm$0.30  & 5.40*         &  175  &  100  & 2.3$^{+0.4}_{-0.3}$ & -- & 15.0$^{+5.0}_{-4.0}$\\
DHCep-2      & 41$\pm$2.0   & 4.3$\pm$0.20  & 5.50*         &  160  &  100  & 1.6$^{+0.2}_{-0.2}$ & 1.0$^{+1.2}_{-1.0}$ & 13.0$^{+1.5}_{-4.0}$\\
V382Cyg-1    & 37$\pm$2.0   & 3.8$\pm$0.15  & 5.17$\pm$0.10 &  260  &  0*  & 2.5$^{+0.5}_{-0.5}$ & 0.7$^{+0.7}_{-0.7}$  & 3.2$^{+1.6}_{-1.6}$  \\
V382Cyg-2    & 38$\pm$3.0   & 3.8$\pm$0.20  & 5.15$\pm$0.14 &  240  &  0*  & 9.0$^{+3.6}_{-3.0}$ & 1.5$^{+0.3}_{-0.4}$  & 6.6$^{+3.0}_{-2.8}$ \\
YCyg-1       & 33$\pm$1.5   & 4.1$\pm$0.20  & 4.57$\pm$0.08 &  140  &  0*  & 1.5$^{+0.5}_{-0.4}$ & 0.5$^{+0.3}_{-0.3}$  & 3.7$^{+1.0}_{-1.0}$ \\
YCyg-2       & 34$\pm$2.0   & 4.2$\pm$0.20  & 4.49$\pm$0.10 &  160  &  0*  & 3.0$^{+2.0}_{-1.1}$ & 0.8$^{+0.6}_{-0.4}$  & 5.0$^{+3.0}_{-1.8}$ \\
V478Cyg-1    & 32$\pm$2.0   & 3.9$\pm$0.20  & 4.74$\pm$0.11 &  120  &  0* & 2.0$^{+0.6}_{-1.5}$ & 0.8$^{+0.8}_{-0.3}$ & 3.4$^{+1.5}_{-1.2}$ \\
V478Cyg-2    & 31$\pm$3.0   & 3.7$\pm$0.30  & 4.78$\pm$0.17 &  120  &  0* & 1.0$^{+0.5}_{-0.5}$ & 0.35$^{+0.7}_{-0.35}$ & 2.6$^{+1.6}_{-1.2}$ \\
XZCep-1      & 28$\pm$1.0   & 3.4$\pm$0.15  & 5.05$\pm$0.06 &  230  &  0*  & $<$2.0 & 0.6$^{+0.6}_{-0.6}$  & 1.0$^{+1.7}_{-0.5}$ \\
XZCep-2      & 24$\pm$3.0   & 3.1$\pm$0.30  & 4.79$\pm$0.22 &  110  &  0*  & 0.3$^{+0.2}_{-0.2}$ & 2.8$^{+3.4}_{-1.3}$ & 0.5$^{+0.6}_{-0.5}$ \\
AHCep-1      & 31$\pm$3.0   & 4.1$\pm$0.30  & 4.41$\pm$0.17 &  200  &  0*  & 1.0$^{+0.9}_{-0.9}$ & $<$0.2 & 2.8$^{+1.2}_{-1.5}$\\
AHCep-2      & 29$\pm$4.0   & 4.2$\pm$0.30  & 4.15$\pm$0.24 &  170  &  0*  & $<$2.0 & 0.35$^{+0.35}_{-0.25}$ & 2.3$^{+0.9}_{-0.9}$\\
\hline
\end{tabular}
\tablefoot{A * symbol indicated adopted values from \citet{msh05}. Luminosities are computed from \teff\ and $R_{\mathrm{mean}}$ (Table \ref{tab_phoebe}).}
\end{center}
\end{table*}

\subsection{Photometry}
\label{s_phot}

Except for DH\,Cep, the five other systems display eclipses. We thus used the PHOEBE (PHysics Of Eclipsing BinariEs, \citealt{prsa05}, v0.31a) software to model the \emph{Hipparcos} lightcurves \citep{hip97}. For their analysis, we kept the parameters of the orbit fixed at their values obtained through the spectroscopic analysis (Section\,\ref{orbit}). PHOEBE allows to model the lightcurve and the RV curves of an object at the same time. This software is based on Wilson \& Devinney’s code \citep{wilson71} and uses Nelder \& Mead’s Simplex fitting method to adjust all the input parameters to find the best fit to the lightcurve. 
In the search for the best-fit model, we include in the computation of the lightcurve the reflection effect. The latter is reported as the brightening of one star due to irradiation by its companion and has a non-negligible impact on very luminous systems when the two component orbit very close to each other. The reflection effect can affect the curvature of the lightcurve in zones outside the eclipse. Reflection is treated according to the theory of \citet{wilson90} in the code PHOEBE.
We also fixed the effective temperatures to the values obtained in the previous section. The photometric parameters derived by PHOEBE are given in Table\,\ref{tab_phoebe}. The error bars are computed by fixing one parameter at a time and we allow the others to vary to reach the minimum of the $\chi^2$ corresponding to a 68.3\% confidence level (1-$\sigma$).

\begin{sidewaystable*}
\begin{center}
\caption{Photometric parameters obtained with PHOEBE. The 1-$\sigma$ errors are given. }  
\label{tab_phoebe}  
\begin{tabular}{l c c c c c c c c c c}  
  \hline\hline
  & \multicolumn{2}{c}{AH\,Cep} &  \multicolumn{2}{c}{XZ\,Cep} & \multicolumn{2}{c}{V382\,Cyg} &  \multicolumn{2}{c}{V478\,Cyg} &  \multicolumn{2}{c}{Y\,Cyg}\\
  & Primary      & Secondary    & Primary      & Secondary     & Primary      & Secondary      & Primary      & Secondary       & Primary     & Secondary \\
  \hline
  $i$ [$\degr$] & \multicolumn{2}{c}{$73 \pm 2$}                     & \multicolumn{2}{c}{$80 \pm 2$}          & \multicolumn{2}{c}{$85 \pm 1$}       & \multicolumn{2}{c}{$78 \pm 1$}         & \multicolumn{2}{c}{$88 \pm 2$} \\
  $R_{\mathrm{mean}}$ [\rsun] & $ 5.6 \pm 0.1 $ & $4.7 \pm 0.1$      & $14.2 \pm 0.1 $ &  $14.2 \pm 0.1 $      & $9.4  \pm 0.2$ &  $8.7 \pm 0.2$      & $7.6 \pm 0.1 $ &  $8.5 \pm 0.1 $       & $5.9 \pm 0.1 $ &  $5.1 \pm 0.1 $ \\
  $R_{\mathrm{pole}}$ [\rsun] & $ 5.5 \pm 0.1 $ & $4.6 \pm 0.1 $     & $13.7 \pm 0.1 $ &  $12.8 \pm 0.1 $      & $8.8  \pm 0.2$ &  $8.0 \pm 0.2$      & $7.4 \pm 0.1 $ &  $8.2 \pm 0.1 $       & $5.9 \pm 0.1 $ &  $5.0 \pm 0.1 $ \\
  $R_{\mathrm{point}}$ [\rsun]& $ 5.9 \pm 0.1 $ & $4.9 \pm 0.1 $     & $15.3 \pm 0.1 $ &  $ --  \pm 0.1 $      & $ --  \pm 0.2$ &  $ -- \pm 0.2$      & $8.0 \pm 0.1 $ &  $9.3 \pm 0.1 $       & $6.1 \pm 0.1 $ &  $5.2 \pm 0.1 $   \\
  $R_{\mathrm{side}}$ [\rsun] & $ 5.6 \pm 0.1 $ & $4.7 \pm 0.1 $     & $14.2 \pm 0.1 $ &  $13.6 \pm 0.1 $      & $9.2 \pm 0.2$ &  $8.5 \pm 0.2$      & $7.6 \pm 0.1 $ &  $8.5 \pm 0.1 $       & $5.9 \pm 0.1 $ &  $5.1 \pm 0.1 $   \\
  $R_{\mathrm{back}}$ [\rsun] & $ 5.8 \pm 0.1 $ & $4.8 \pm 0.1 $     & $14.8 \pm 0.1 $ &  $16.4 \pm 0.1 $      & $9.9 \pm 0.2$ &  $9.5 \pm 0.2$      & $7.8 \pm 0.1 $ &  $8.9 \pm 0.1 $       & $6.0 \pm 0.1 $ &  $5.1 \pm 0.1 $   \\
  $M_{\mathrm{bol}}$ [mag]    & $-6.26 \pm 0.05 $ & $-5.59 \pm 0.05$ & $-7.84 \pm 0.07 $ & $-7.16 \pm 0.07 $   & $-8.30 \pm 0.08$ & $-8.03 \pm 0.08 $ & $-7.05 \pm 0.06 $ & $-7.16 \pm 0.06$   & $-6.65 \pm 0.04 $ & $-6.44 \pm 0.04 $ \\
  $l_p/l_s$                   &  \multicolumn{2}{c}{$1.82\pm0.18$}   &  \multicolumn{2}{c}{$1.86\pm0.28$}      &  \multicolumn{2}{c}{$1.05\pm0.26$}   &  \multicolumn{2}{c}{$0.95 \pm 0.10$}       &  \multicolumn{2}{c}{$1.23 \pm 0.12$}\\
  \hline
\end{tabular}
\tablefoot{$R_{\mathrm{pole}}$ is the radius toward the pole, $R_{\mathrm{point}}$ is the radius measured toward the other component, $R_{\mathrm{side}}$ is the radius measured toward the side, $R_{\mathrm{back}}$ toward the Lagrangian point $L_2$ and $R_{\mathrm{mean}}$ is the geometrical mean between $R_{\mathrm{pole}}$, $R_{\mathrm{side}}$ and $R_{\mathrm{back}}$. The bolometric albedos (coefficients of reprocessing of the emission of a companion by "reflection") and the gravity darkening coefficients are both set to 1.0 for the different components. $l_p/l_s$ is the ratio of the primary's to the secondary's luminosity.}
\end{center}
\end{sidewaystable*}

\section{Discussion}
\label{s_disc}

\subsection{Comparison to previous analysis}
\label{s_comp}

\begin{table*}
\begin{center}
\caption{Mass and radius from the literature.} \label{tab_comp}
\begin{tabular}{lrrrrl}
\hline
Star         & M$_p$  &  M$_s$  & R$_p$   & R$_s$  & Reference\\    
             & [\msun]& [\msun] & [\rsun]& [\rsun]\\
\hline
V478~Cyg      & 15.3$\pm$0.8   & 14.6$\pm$0.8   & 7.6$\pm$0.1     & 8.5$\pm$0.1     & this work \\
              & 16.6$\pm$0.9   & 16.3$\pm$0.9   & 7.43$\pm$0.12   & 7.43$\pm$0.12   & \citet{ph91} \\
Y~Cyg         & 16.9$\pm$0.5   & 16.0$\pm$0.5   & 5.9$\pm$0.1     & 5.1$\pm$0.1     & this work \\
              & 16.8$\pm$0.5   & 17.7$\pm$0.4   & --              & --              & \citet{burk97} \\
              & 17.72$\pm$0.35 & 17.73$\pm$0.30 & 5.785$\pm$0.091 & 5.816$\pm$0.063 & \citet{harm14} \\
              & 17.57$\pm$0.27 & 17.04$\pm$0.26 & 5.93$\pm$0.07   & 5.78$\pm$0.07   & \citet{ss94b}\\
V382~Cyg      & 26.1$\pm$0.4   & 19.0$\pm$0.3   & 9.4$\pm$0.2     & 8.7$\pm$0.2     & this work \\
              & 32.6$\pm$1.8   & 22.9$\pm$1.3   & 8.8$\pm$0.6     & 7.4$\pm$0.6     & \citet{ph91} \\
              & 29.7$\pm$1.1   & 20.3$\pm$0.9   & --              & --              & \citet{burk97} \\
              & 26.0$\pm$0.7   & 19.3$\pm$0.4   & 9.6$\pm$0.1     & 8.4$\pm$0.1     & \citet{harries97} \\
XZ~Cep        & 18.7$\pm$1.3   &  9.3$\pm$0.5   & 14.2$\pm$0.1    & 14.2$\pm$0.1    & this work \\
              & 15.8$\pm$0.4   &  6.4$\pm$0.3   &  7.0$\pm$0.2    & 140.5$\pm$0.2   & \citet{harries97} \\
AH~Cep        & 14.3$\pm$1.0   & 12.6$\pm$0.9   & 5.6$\pm$0.1     & 4.7$\pm$0.1     & this work \\
              & 16.2$^{+6.0}_{2.5}$ & 13.3$^{+5.5}_{2.3}$ & --         & --              & \citet{burk97} \\
              & 18.1$\pm$0.9   & 15.9$\pm$0.8   & 6.7$\pm$0.2     & 6.2$\pm$0.2     & \citet{bell86} \\
DH~Cep$^1$    & 38.4$\pm$2.5     & 33.4$\pm$2.2          & --              & --              & this work \\
              & 34.4$^{+2.8}_{-2.5}$ & 29.8$^{+2.5}_{-2.4}$   & --       & --              & \citet{burk97} \\
              & 32.7$\pm$1.7      & 29.6$\pm$1.6         & --              & --              &  \citet{hild96}\\
\hline
\end{tabular}
\tablefoot{1- For DH~Cep, we assume an inclination of 47$^o \pm$1, according to \citet{ss94}, to obtain the dynamical masses.}
\end{center}
\end{table*}

In this section we compare our results with those available in the literature. Table \ref{tab_comp} gathers the published values of masses and radii for the systems we have analyzed.

V478~Cyg was studied by \citet{ph91}. Within the error bars, their results are in excellent agreement with our findings with the exception of the radius of the secondary which is slightly larger in our study. 

\citet{ph91} also analyzed V382~Cyg. Our radii are marginally consistent with their results, while our mass estimates are lower. The masses of \citet{burk97} are in better agreement with our findings. \citet{burk97} concluded that V382~Cyg was an interacting system since both stars fill their Roche lobe. Similar conclusions were reached by \citet{harries97} who obtained masses and radii in excellent agreement with our results.

The mass and radius of the primary component of Y~Cyg obtained by \citet{burk97}, \citet{harm14} and \citet{ss94b} are consistent with our values. On the contrary, we find a less massive and more compact secondary. We note that this system is the only one for which the phase coverage of our observations does not sample the maximum separation (Fig.\ \ref{fig:orbital}). This probably affects our results and explains the differences with previous studies.  
The effective temperature and surface gravity derived by \citet{ss94b} -- \teff\ = 34500 (34200) K and \logg\ = 4.16 (4.18) for the primary (secondary) -- are similar to our determinations, within the error bars. The same conclusion applies to the determination of the inclination of the system ($i$=86$^o$37$\pm$0$^o$12 for Simon \& Sturm, $i$=88$^o$0$\pm$2.0 for us).

The dynamical mass determination of \citet{burk97} for AH~Cep is consistent with our results. Our masses are $\sim$ 25$\%$ lower and our radii about 20$\%$ smaller than those of \citet{bell86}. 

The inclination of DH~Cep was estimated by \citet{ss94}: they found $i$=47$\pm$1$^o$. Adopting this value, we obtain masses of 38.4$\pm$2.5 and 33.4$\pm$2.2 \msun\ for the primary and secondary. Within the error bars these values are in agreement with those of \citet{burk97} and \citet{hild96}.\\
Our effective temperatures are consistent with those of \citet{ss94} -- \teff\ = 44000 (43000) K for the primary (secondary) --  but we find surface gravities larger by $\sim$0.3 dex (Simon \& Sturm report \logg\ = 3.95 (4.03) for the primary (secondary)). This likely reflects the difficulty to correctly disentangle the individual profiles of Balmer lines that are the main gravity indicators.

Finally, XZ~Cep was studied by \citet{harries97} who obtained masses lower by $\sim$3 \msun\ compared to our determinations. On the contrary their radii are slightly larger. These differences are probably due to the contact nature of the system (see below).

\subsection{Evolutionary status}
\label{s_evol}

\begin{figure*}[]
\centering
\includegraphics[width=0.47\textwidth]{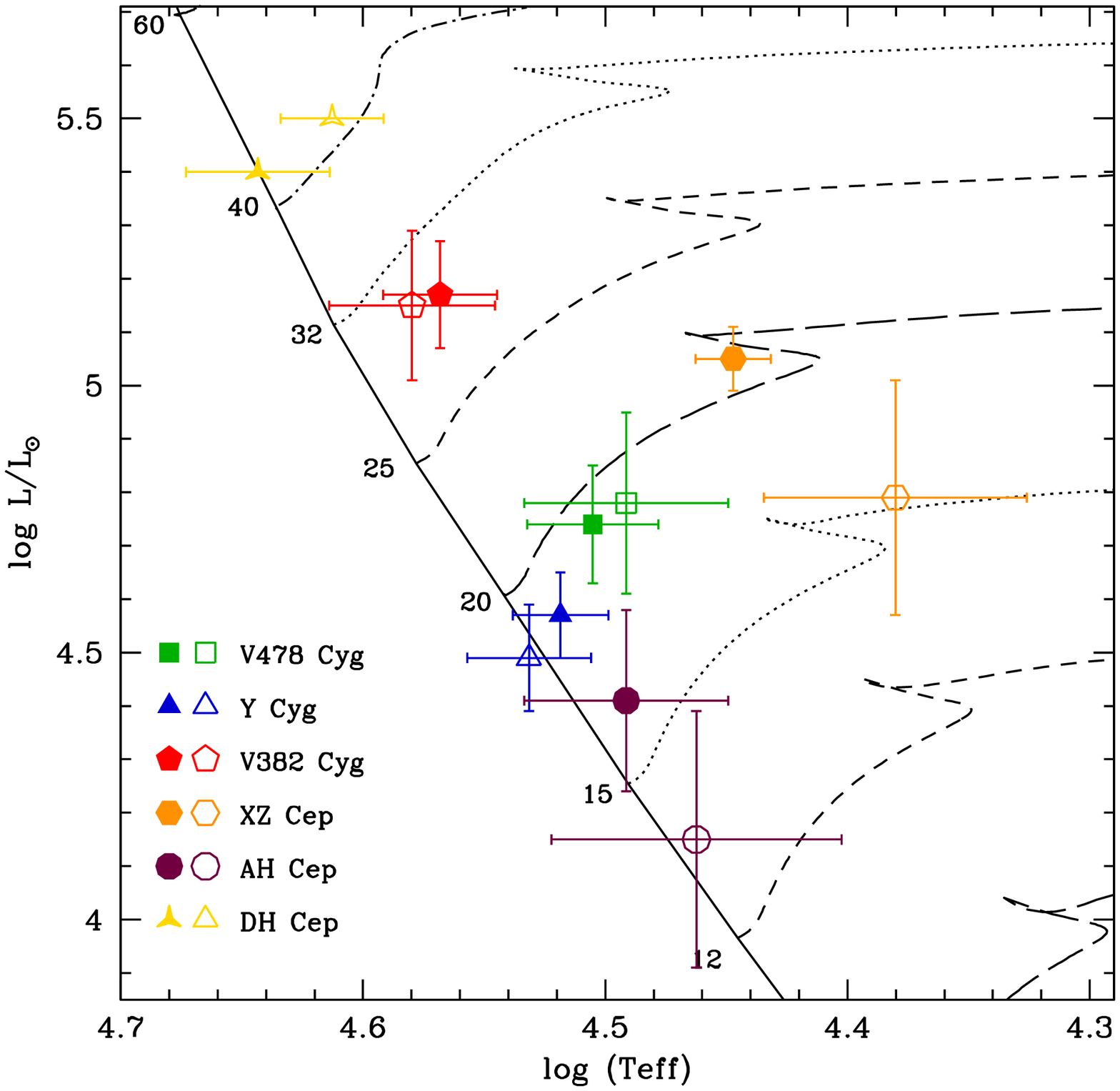}
\includegraphics[width=0.47\textwidth]{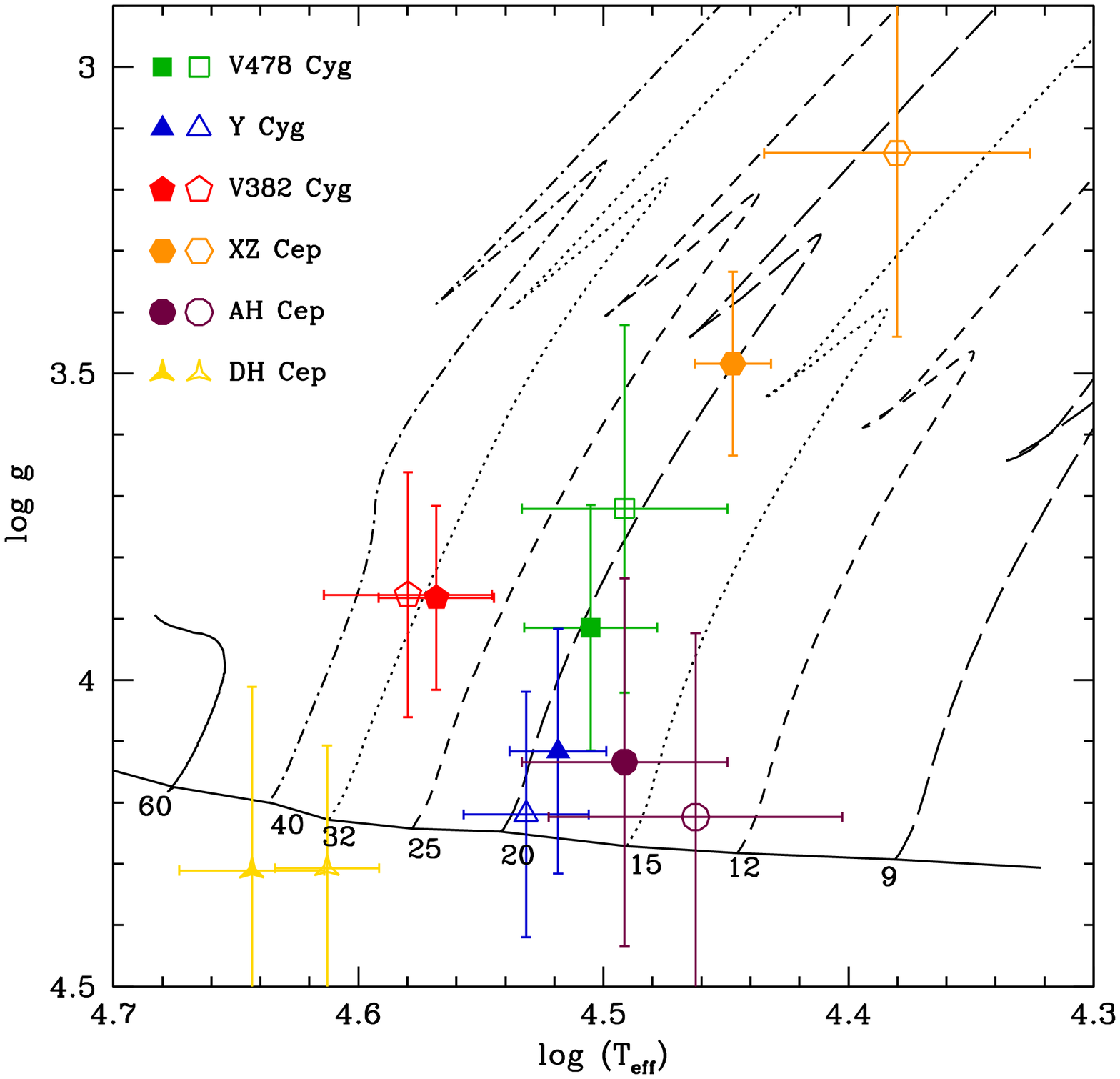}
\caption{\textit{Left:} Hertzsprung-Russell diagram. \textit{Right:} Surface gravity as a function of effective temperature for the sample stars. Evolutionary tracks are from \citet{ek12}. Filled (open) symbols refer to the primary (secondary) of each binary system.}
\label{fig_hr}
\end{figure*}

Fig.\ \ref{fig_hr} shows the classical Hertzsprung-Russell (HR) diagram and the \logg - \teff\ diagram. All systems but XZ~Cep have both of their components on the main sequence as defined by single-star evolutionary tracks. XZ~Cep is the most evolved system of our sample. For almost all stars the same track can explain the position of stars in both the HR and \logg - \teff\ diagrams. The secondary component of XZ~Cep may be the only exception since its luminosity places it close to the 15 \msun\ track in the HR diagram while a mass of 20 \msun\ better reproduces its position in the \logg - \teff\ diagram. But we also note that within the error bars the 15 and 20 \msun\ tracks are relevant in both diagrams.

Table \ref{tab_RL} provides an estimate of the Roche lobe size for each component of the five eclipsing systems. This estimate was made using equation 2 of \citet{eggleton83}. In the third column of Table~\ref{tab_RL} we compare the size of the Roche lobe to the stellar radius in the direction of the companion ($R_{\mathrm{point}}$) when possible, or to the mean radius when $R_{\mathrm{point}}$ is not available (see Table \ref{tab_phoebe}). V478~Cyg, Y~Cyg and AH~Cep are detached systems. The latter two have the largest ratios R$_{\rm L}$/$R_{\mathrm{point}}$ which is consistent with their position very close to the ZAMS. V478~Cyg, which is more evolved, is on the verge of filling its Roche lobe (secondary star). V382~Cyg and XZ~Cep are contact systems, in agreement with the findings of \citet{harries97}. Given the above remarks regarding the position on the HR and \logg-\teff\ diagrams, XZ~Cep has most likely experienced mass transfer and certainly follows an evolution different from that of single stars (see also Sect.\ \ref{s_ab}). The properties of V382~Cyg in both evolutionary diagrams do not indicate deviation from single star evolution as strong as for XZ~Cep: V382~Cyg is thus probably only in an early contact phase so that deviation from single-star evolution has not appeared yet. 

It is interesting to compare Y~Cyg and V478~Cyg in the light of Table~\ref{tab_RL}: both stars have a mass ratio close to 1.0, dynamical masses around 15 \msun, and similar orbital periods (just below 3 days). According to Fig.\ \ref{fig_hr} Y~Cyg is located closer to the ZAMS and is less evolved than V478~Cyg which lies in the middle of the main sequence. This is reflected in Table \ref{tab_RL} where the former system has its components filling a smaller fraction of its Roche lobe than the latter. We conclude that for an orbital period of about 3 days and stars with similar initial masses of about 15 \msun\ contact does not happen before the final portion of the main sequence. 

\begin{table}
\begin{center}
\caption{Size of the Roche lobe (R$_{\rm L}$) and stellar radius towards the companion ($R_{\mathrm{point}}$), all in units of solar radius.} \label{tab_RL}
\begin{tabular}{lcccccccc}
\hline
Star         & R$_{\rm L}$  & $R_{\mathrm{point}}$   &  R$_{\rm L}$/$R_{\mathrm{point}}$   \\    
             &     &         &          \\
\hline
V478Cyg-1    & 10.1  & 8.0$\pm 0.1$       & 1.26$\pm 0.1$ \\
V478Cyg-2    & 9.9   & 9.3$\pm 0.1$       & 1.06$\pm 0.1$ \\
YCyg-1       & 10.9  & 6.1$\pm 0.1$       & 1.78$\pm 0.1$ \\
YCyg-2       & 10.4  & 5.2$\pm 0.1$       & 2.00$\pm 0.1$ \\
V382Cyg-1    & 9.3   & 9.4$\pm 0.2$ $^1$  & 0.99$\pm 0.2$ \\
V382Cyg-2    & 8.0   & 8.7$\pm 0.2$ $^1$  & 0.92$\pm 0.2$ \\
XZCep-1      & 16.7  & 15.3$\pm 0.1$      & 1.09$\pm 0.1$ \\
XZCep-2      & 12.1  & 14.2$\pm 0.1$ $^1$ & 0.85$\pm 0.1$ \\
AHCep-1      & 7.2   & 5.9$\pm 0.1$       & 1.22$\pm 0.1$ \\
AHCep-2      & 6.8   & 4.9$\pm 0.1$       & 1.39$\pm 0.1$ \\
\hline
\end{tabular}
\tablefoot{1- $R_{\mathrm{mean}}$ is used instead of $R_{\mathrm{point}}$ since the latter is not available.}
\end{center}
\end{table}

\subsection{Surface abundances}
\label{s_ab}

\begin{figure}[]
\centering
\includegraphics[width=0.47\textwidth]{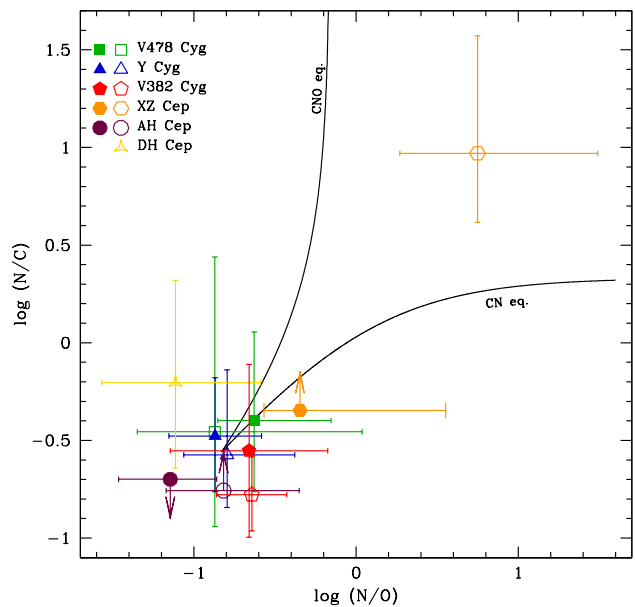}
\caption{Logarithm of N/C as a function of Logarithm N/O (left). Solid lines indicate values for complete or partial CN(O) cycle. Symbols have the same meaning as in Fig.\ \ref{fig_hr}.}
\label{fig_ab}
\end{figure}

Fig. \ \ref{fig_ab} shows that most of the sample stars are barely evolved in terms of surface CNO abundances. Almost all stars have C, N and O abundance patterns consistent with no processing. The only exception is the secondary of XZ~Cep which is significantly enriched. The ratios N/C and N/O are consistent with CNO burning.

\begin{figure*}[]
\centering
\includegraphics[width=0.32\textwidth]{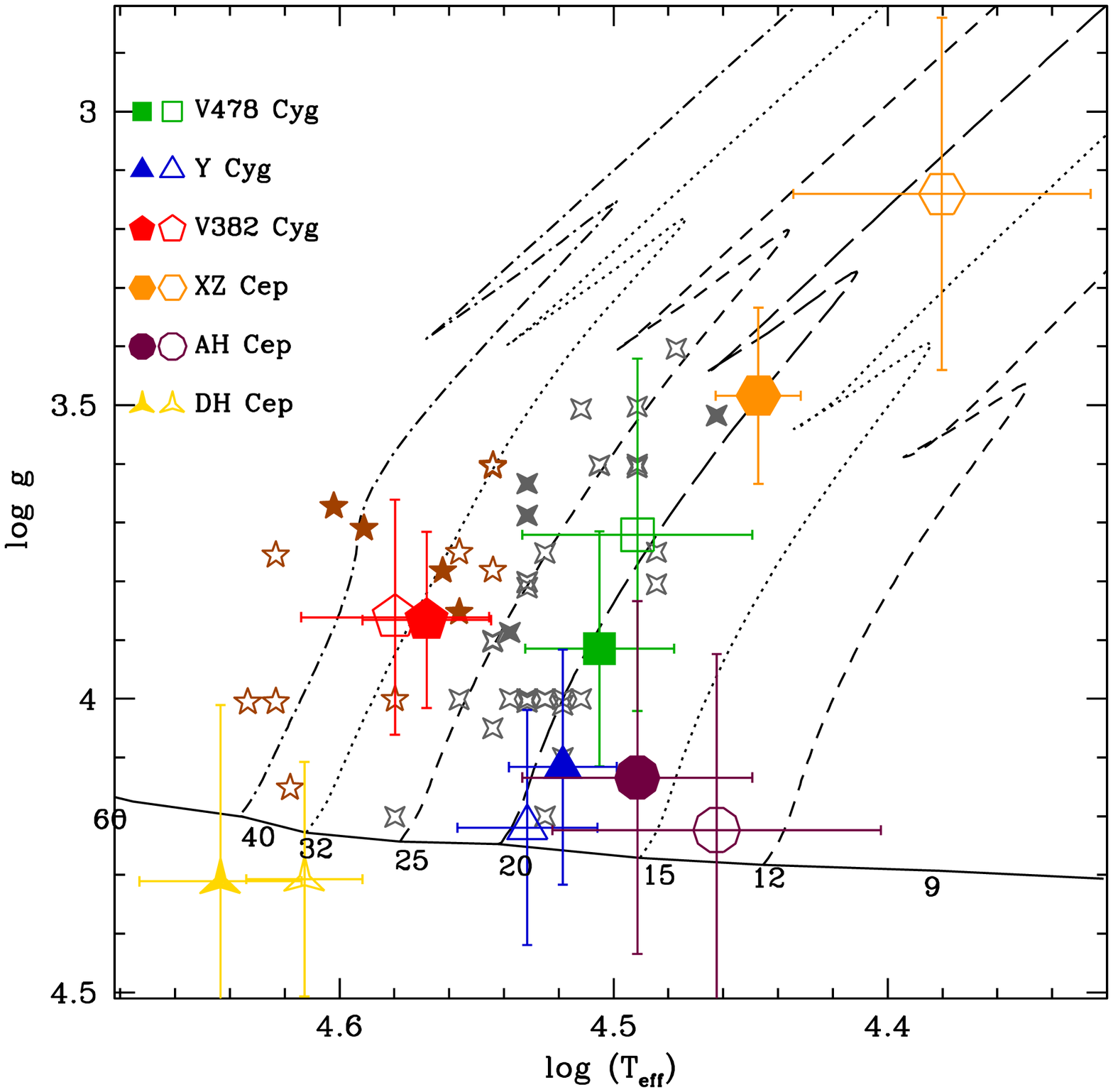}
\includegraphics[width=0.32\textwidth]{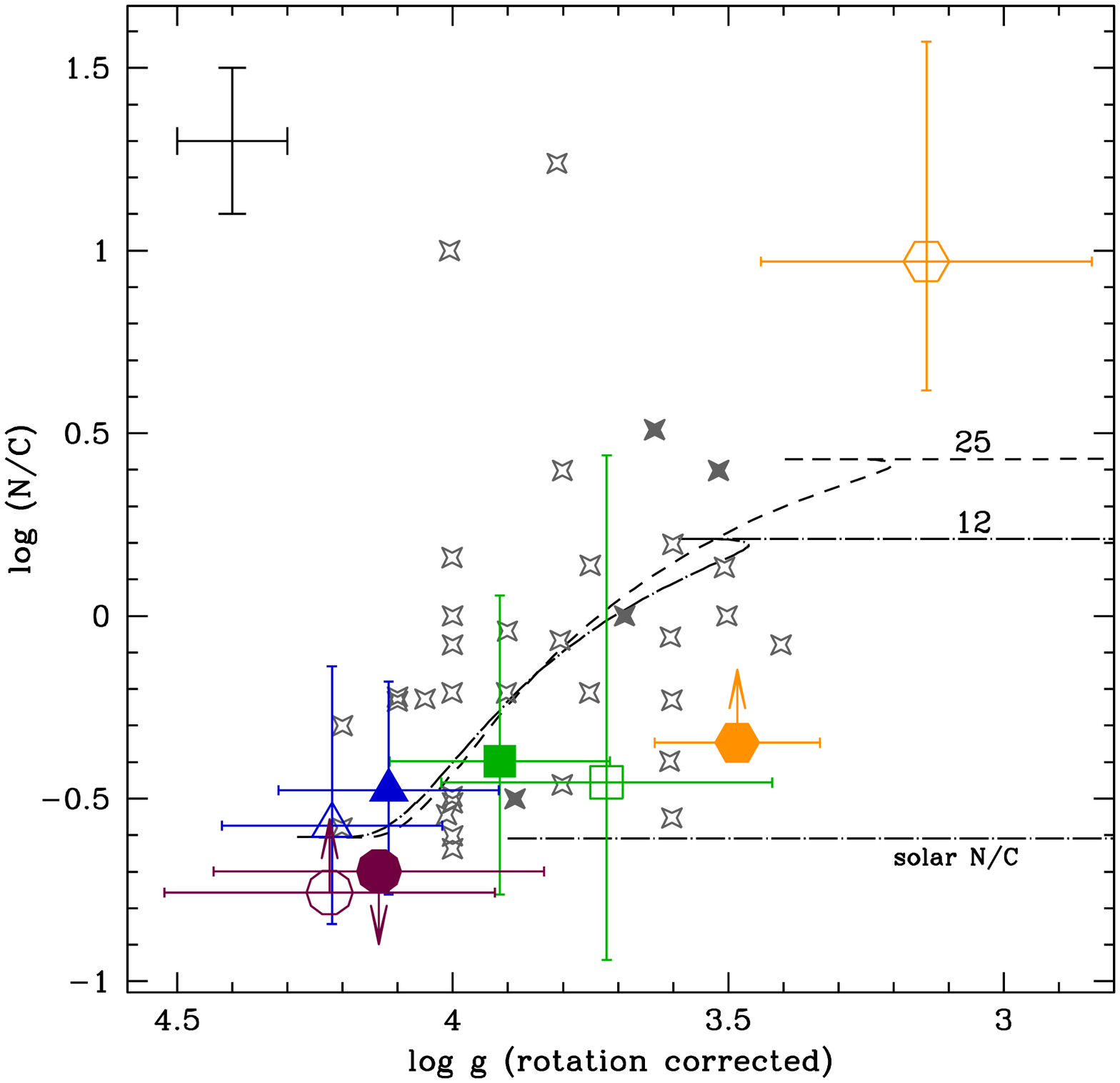}
\includegraphics[width=0.32\textwidth]{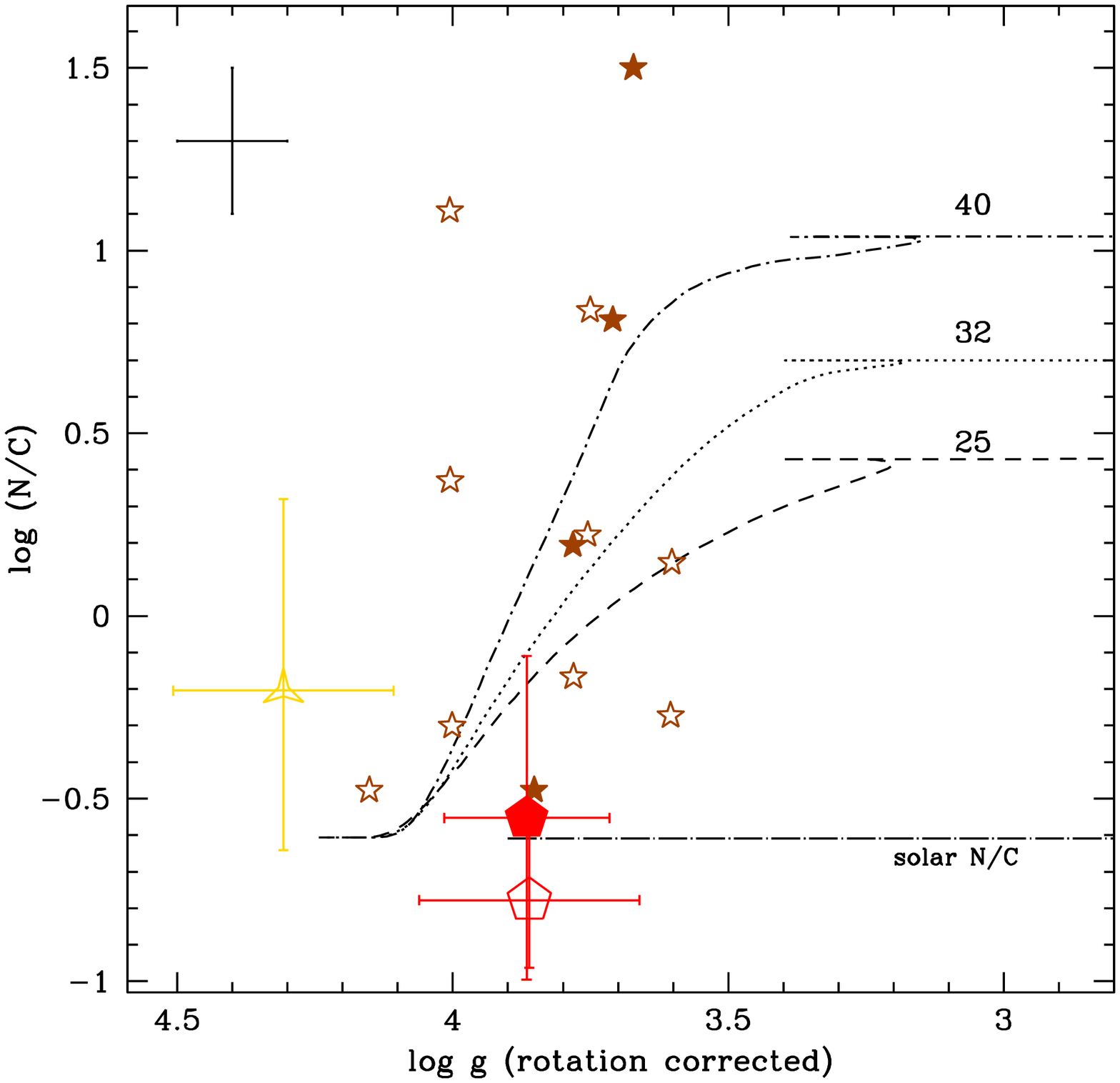}
\caption{\textit{Left:} \logg - \teff\ (left) diagram for the sample stars together with comparison single stars from \citet{mimesO}. In the comparison sample, grey asterisks (brown stars) refer to stars less (more) massive than $\sim$28 \msun. Filled symbols (stars and asterisks) refer to comparison stars with \vsini\ $>$ 120 \kms. For clarity only a representative error bar for the comparison sample is shown. The evolutionary models including rotation of \citet{ek12} are shown. \textit{Middle and right:} $\log$(N/C) - \logg\ diagram for stars with masses lower than 28 \msun\ (middle) and higher than 28 \msun (right). In all panels big symbols have the same meaning as in Fig.\ \ref{fig_hr}.}
\label{fig_abcomp}
\end{figure*}

Fig.\ \ref{fig_abcomp} shows a comparison of the sample stars with presumably single stars analyzed by \citet{mimesO}. We chose this study as a reference for comparison since it is based on data of similar quality and the method employed to determine the stellar parameters and surface abundances is the same as the one we used in the present study. The left panel of Fig.\ \ref{fig_abcomp} shows the \logg - \teff\ diagram. The comparison stars are separated in two sub-samples depending on their initial mass (above or below $\sim$28 \msun) according to the tracks of \citet{ek12}. In the middle  panel, the surface abundances of V478~Cyg, Y~Cyg, XZ~Cep and AH~Cep can be compared to the single stars with masses below 28 \msun. V382~Cyg and DH~Cep are compared to more massive single stars in the right panel. We see that in both mass ranges there is no difference between the binary and single star samples: they cover the same area of the log(N/C) -- \logg\ diagram. As above, the only possible difference is the secondary star of XZ~Cep. Although there is no direct comparison single stars close to it, we see that its N/C ratio corresponds to predictions of single star models with initial masses above 25 \msun\ (and close to 40 \msun), while its estimated mass is in the range 8-20 \msun\ (see Sect. \ref{s_mass}). It is thus likely that in this evolved system binarity has affected the surface chemical composition. For the other five systems, there is no evidence for such an effect, even for V382~Cyg in which both components (barely) fill their Roche lobe. 
In Fig.\ \ref{fig_abcomp} we have also separated comparison stars according to whether they have \vsini\ higher than 120 \kms\ or not. The high-\vsini\ sample encompass the range of projected rotational velocities of the binary components. Here again, there is no clear distinction between binaries and single stars.  

We thus conclude that in the binary sample we have studied, detached systems do not show clear differences in their surface abundances compared to single stars. For interacting systems, large N/C may be encountered (XZ~Cep) or not (V382~Cyg). A certain time may be required after the beginning of the mass transfer in order to see its effects on surface abundances. 
 
\vspace{0.5cm}

\citet{garland17} studied a sample of 33 B-type binary systems in the Large Magellanic Cloud, all of them located on the main sequence. They did not perform a disentangling of the individual spectra but determined the stellar parameters of the primaries with atmosphere models. They used various assumptions regarding contamination of their spectra by the secondaries to infer the impact on the derived parameters. They estimated nitrogen surface abundances and projected rotational velocities and concluded that primary stars of their binary systems do not show differences in terms of surface enrichment compared to B-type single stars. They argue that they may have evolved effectively as single stars with low rotational velocities. 

\citet{mahy17b} determined the stellar parameters and surface abundances of the O7.5If+O9I(f) system HD~166734. Their analysis indicates that this is a detached system and that the degree of chemical processing is not different from what is observed in single O supergiants.

\begin{figure}[]
\centering
\includegraphics[width=0.47\textwidth]{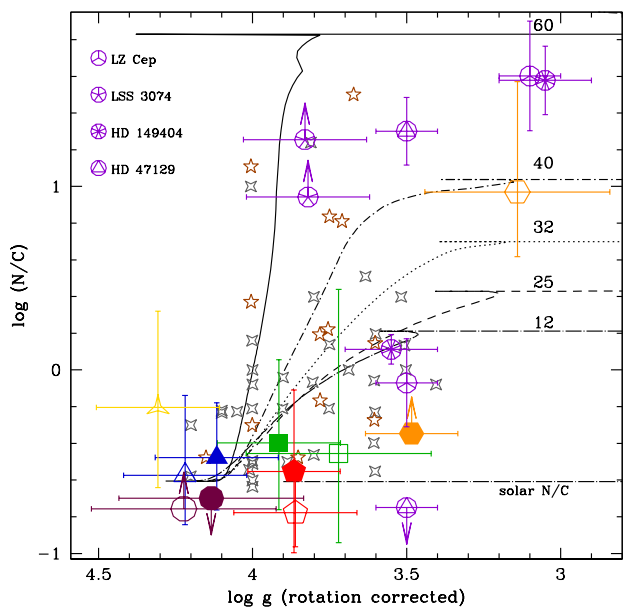}
\caption{Same as middle and right panels of Fig.\ \ref{fig_abcomp} but with the evolved systems HD~47129 \citep{linder08}, LZ~Cep \citep{mahy11}, HD~149404 \citep{raucq16} and LSS~3074 \citep{raucq17} added. Big symbols have the same meaning as in Fig.\ \ref{fig_hr}.}
\label{fig_abcompBin}
\end{figure}

\citet{mahy11} analyzed the properties of the two components of the massive binary LZ~Cep. They found that the secondary is extremely nitrogen- and helium-rich, while being carbon- and oxygen-poor. Quantitatively, $\log$(N/C)=1.6 which, together with a \logg\ of 3.1, places the star in the upper right corner of Fig.\ \ref{fig_abcompBin} (the present-day dynamical mass of the secondary is $\sim$ 6 \msun). On the other hand the primary is barely chemically evolved. \citet{mahy11} concluded that the surface abundances could be explained by an inefficient mass transfer from the secondary (which was initially the primary) towards the secondary. The dynamical masses obtained by Mahy et al.\ for LZ~Cep ($\sim$ 16 and $\sim$6 \msun\ for the primary and secondary) are significantly lower than the evolutionary masses (25.3 and 18.0 \msun) determined using single star evolutionary tracks. This is another indication that mass transfer occurred in that system according to \citet{mahy11}. In any case, the degree of chemical enrichment is higher than expected for single stars of masses 5 to 25 \msun -- it would be consistent with the enrichment of a $\sim$50 \msun\ star.

The surface abundances of the components of the system HD149404 are similar to that of LZ~Cep according to \citet{raucq16} -- see also Fig.\ \ref{fig_abcompBin}. Since HD~149404 is not an eclipsing binary, there is no determination of its dynamical mass. The evolutionary masses reported by \citet{raucq16} are of the order 30 \msun\  assuming a standard initial rotation rate. As for LZ~Cep, the secondary is thus highly enriched, as confirmed by its position in Fig.\ \ref{fig_abcompBin}: the observed N/C ratio is larger by $\sim$0.8 dex compared to the prediction of a 32 \msun\ model. Given these abundances and the asynchronous rotation of the system, \citet{raucq16} suggested that HD149404 is in a post-Roche lobe overflow state. Such an interaction may have revealed internal layers of the secondary star, explaining the rather strong enrichment. 

The massive evolved system LSS~3074 was studied by \citet{raucq17}. The dynamical masses of the components are 14.8 and 17.2 \msun\ for the primary and secondary respectively. These values (together with the estimated radii) are low for stars with spectral type O4f and O6-7(f). Raucq et al.\ determined lower limits of 0.94 (1.26) for the primary (secondary) $\log$(N/C) ratios. The primary is also helium-rich (He/H=0.25). The surface gravities of both components being close to 3.8, the stars are located in the upper part of Fig.\ \ref{fig_abcompBin}. \citet{raucq17} indicate that LSS~3074 is likely an overcontact system which experienced Roche lobe overflow. Their simulations favour a relatively massive primary (M=35 \msun). In that case, the N/C ratio determined for that star is high compared to predictions of evolutionary models (see Fig.\ \ref{fig_abcompBin}). Compared to the two systems described above, it is worth noting that in the case of LSS~3074 both components show a high N/C ratio. 

Both components of Plaskett's star (HD~47129) have initial masses above 40 \msun\ according to \citet{linder08}. The primary has a high N/C ratio, while the secondary is surprisingly chemically unevolved. The secondary has a high projected rotational velocity, which prompted Linder et al.\ to argue that this system experienced a recent mass transfer in which a small fraction of the mass removed from the primary was accreted by the secondary. This could explain the high rotation and absence of nitrogen enrichment / carbon depletion of the secondary. In Fig.\ \ref{fig_abcompBin} the primary is located in the upper part at a position consistent with high mass single stars. We note that a strong magnetic field is present at the surface of the secondary \citet{gru13}, which may have affected the interaction history and surface abundances of that system \citep[e.g.][]{meynet11}.

From the above discussion, and keeping in mind that samples of binary stars for which surface abundances are available remain small, the following trend tends to emerge:

\begin{itemize}

\item systems for which no evidence of mass transfer exist do not show surface enrichment significantly different from single stars, at least within the uncertainties on the derived surface abundances. Said differently, the additional role of tides on mixing in detached systems has a limited effect on surface abundances compared to mixing in single stars.

\item on the contrary in three systems which experienced mass transfer (HD~47129, LZ~Cep and HD~149404), the mass donor has a high N/C ratio, even higher than what is observed in single stars of the same initial mass and rotation rate for two systems (LZ~Cep and HD~149404). Our results for XZ~Cep follow the same trend: the secondary fills its Roche lobe and has a larger N/C than expected for its mass. In these systems mass transfers probably removed the external layers of the donor. This process exposes at the surface internal regions more chemically mixed (an effect similar to stellar winds for Wolf-Rayet stars). The mass gainer do not show peculiar surface abundances compared to single stars. This latter result is consistent with predictions of \citet{langer08} who show that the accreted material is mixed and diluted in the pre-existing envelope. For HD~149404, both the primary and secondary show relatively high N/C ratios, possibly indicating that the mass gainer was already chemically enriched (because of rotational mixing) before interaction. Finally, V382~Cyg just fills its Roche lobe but do not show signs of peculiar N/C.

\end{itemize}
 
Peculiar N/C ratios would thus be observed in the mass donor of post-mass transfer systems, possibly because of the removal of external layers and hence the appearance of internal layers deeply chemically processed at the surface. In absence of mass transfer, surface abundances of binary components would be indistinguishable from those of single stars. The case of the contact system V382~Cyg, for which no peculiar abundances are determined, suggests that the above picture is only partial or that the system was caught at the very beginning of the mass transfer episode, when significant envelope removal did not occur yet. Obviously, this tentative picture needs to be confirmed by analysis of additional evolved systems. Such a task is difficult because the estimates of initial masses in interacting systems are based on assumptions on the efficiency of mass transfer, on the initial orbital parameters, on the initial mass ratio \citep[e.g.][]{wellstein01,demink09,raucq17}. In addition, evolved systems are not spherically symmetric and the use of 1D atmosphere models may be questioned \citep{palate13}.

\subsection{Stellar masses}
\label{s_mass}

In Table~\ref{tab_mass} we present three mass estimates: the dynamical mass (M$_{dyn}$) results from the orbital solution; the evolutionary mass (M$_{evol}$) is estimated from the position in the HR diagram (Fig.\ \ref{fig_hr}, right panel); the spectroscopic mass (M$_{spec}$) is obtained from the surface gravity and mean radius. 

\begin{table}
\begin{center}
\caption{Dynamical, evolutionary and spectroscopic masses.} \label{tab_mass}
\begin{tabular}{lcccccccc}
\hline
Star         & M$_{dyn}$  & M$_{evol}$   &  M$_{spec}$   \\    
             &     &         &          \\
\hline
V478Cyg-1    & 15.3$\pm$0.8  &  19.2$^{+2.7}_{-2.1}$  & 16.8$\pm$10.0 \\
V478Cyg-2    & 14.6$\pm$0.8  &  18.9$^{+3.7}_{-3.3}$  & 13.2$\pm$13.0 \\
YCyg-1       & 16.9$\pm$0.5  &  18.5$^{+1.9}_{-1.2}$  & 16.0$\pm$9.9 \\
YCyg-2       & 16.0$\pm$0.5  &  18.4$^{+1.4}_{-1.4}$  & 15.1$\pm$9.4 \\
V382Cyg-1    & 26.1$\pm$0.4  &  28.4$^{+4.0}_{-3.3}$  & 20.4$\pm$9.4 \\
V382Cyg-2    & 19.0$\pm$0.3  &  29.6$^{+6.0}_{-5.5}$  & 17.4$\pm$10.8 \\
XZCep-1      & 18.7$\pm$1.3  &  20.4$^{+1.1}_{-1.0}$  & 18.5$\pm$8.5 \\
XZCep-2      &  9.3$\pm$0.5  &  15.0$^{+4.6}_{-2.2}$  & 9.3$\pm$9.0 \\
AHCep-1      & 14.3$\pm$1.0  &  16.1$^{+3.3}_{-2.5}$  & 14.4$\pm$14.0 \\
AHCep-2      & 12.6$\pm$0.9  &  13.7$^{+3.2}_{-2.0}$  & 12.8$\pm$12.0 \\
DHCep-1      & 38.4$\pm$2.5$^1$  &  43.6$^{+5.5}_{-5.5}$  & 55.2$\pm$55.0 \\
DHCep-2      & 33.4$\pm$2.2$^1$  &  38.1$^{+6.9}_{-2.3}$  & 91.4$\pm$56.7 \\
\hline
\end{tabular}
\tablefoot{1- For DH~Cep, the dynamical masses assume an inclination of 47$^o \pm$1, according to \citet{ss94}.}
\end{center}
\end{table}

\begin{figure*}[]
\centering
\includegraphics[width=0.47\textwidth]{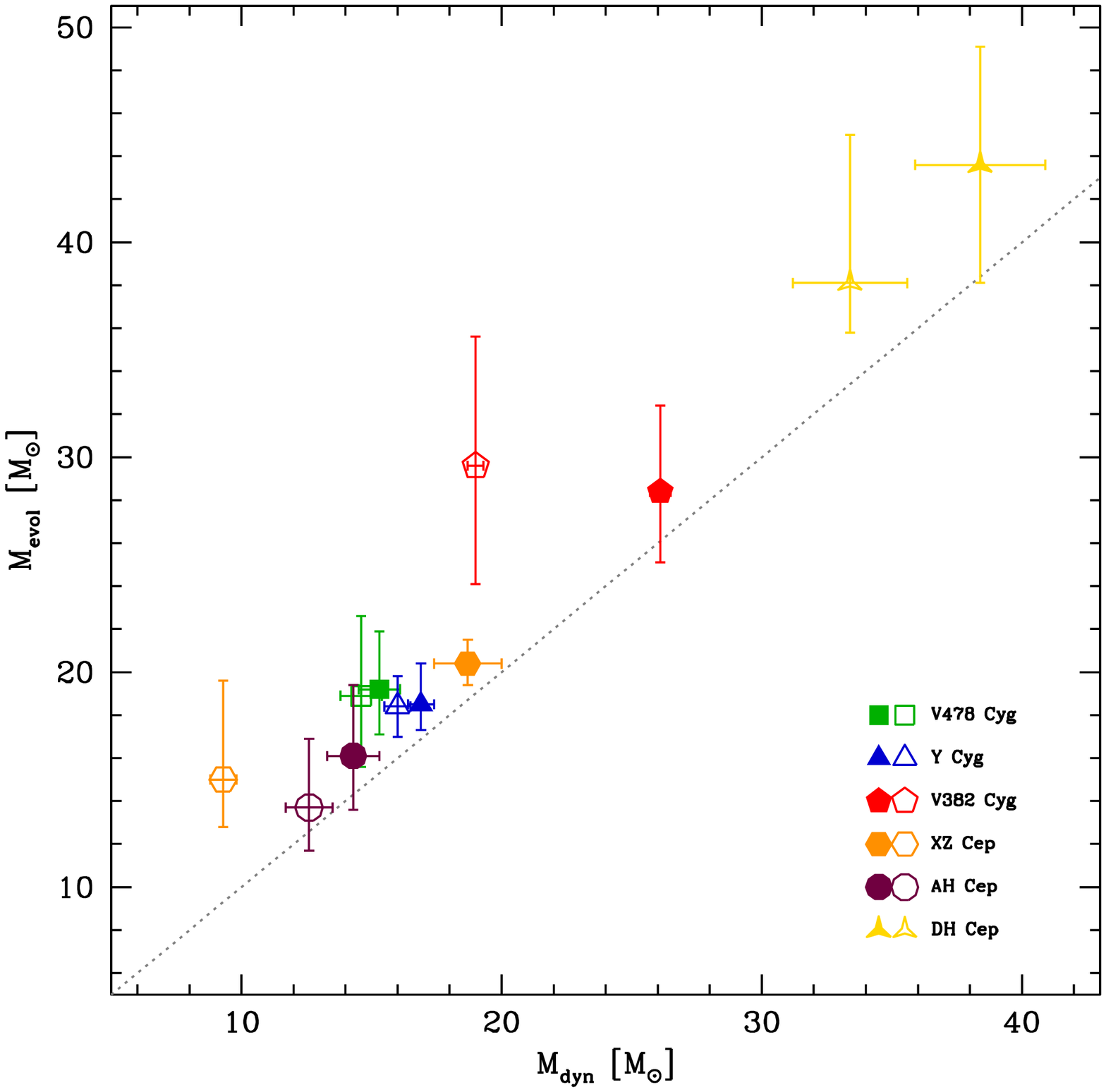}
\includegraphics[width=0.47\textwidth]{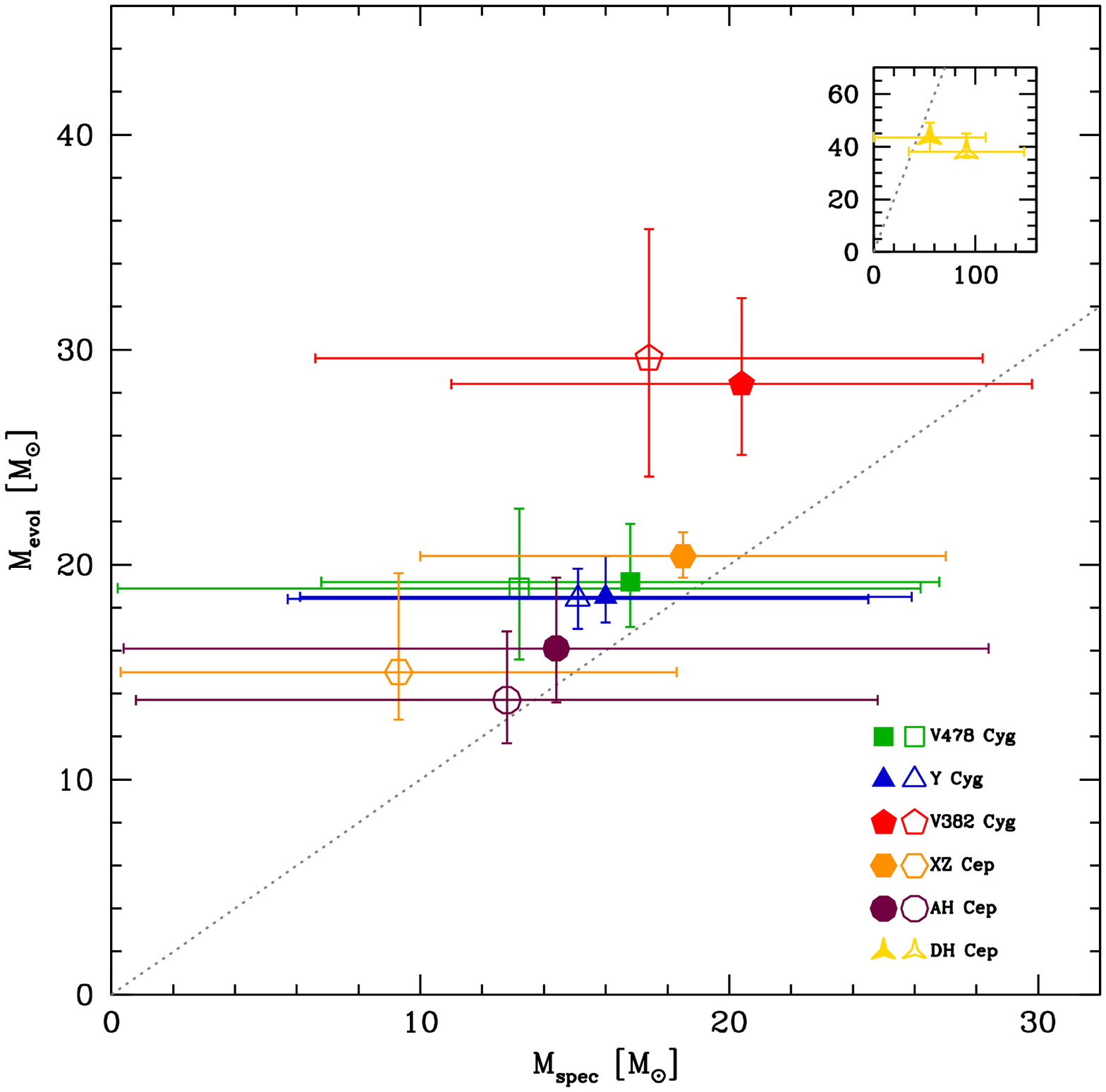}
\caption{Evolutionary mass as a function of dynamical mass (left) and spectroscopic mass (right). Symbols have the same meaning as in Fig.\ \ref{fig_hr}.}
\label{fig_mass}
\end{figure*}

Fig.\ \ref{fig_mass} shows a comparison between these three masses. In the left panel, we see that M$_{dyn}$ and M$_{evol}$ are well correlated. However, there is a systematic upward shift indicating that evolutionary masses tend to be slightly larger (by a few solar masses) than dynamical masses. Quantitatively, the former are 16.5$\pm$10.6\% higher than the latter, with differences ranging between 8 and 38\%. 

\citet{vw10} studied a large sample of binaries and concluded that there was no difference between dynamical and evolutionary masses. The only system in common with our study is V478~Cyg for which \citet{vw10} give M$_{dyn}$=16.6$\pm$9.0 \msun\ (16.3$\pm$9.0 \msun\ for the secondary) and M$_{evol}$=18$^{+4}_{-5}$ \msun. The dynamical masses are in good agreement with our measurements. The evolutionary masses of \citet{vw10} are slightly lower than ours, but consistent within the error bars. Vink \& Weidner relied on the tracks of \citet{mm03} which are less luminous, for a given initial mass, than those of \citet{ek12} that we use. With the 2003 Geneva tracks, one finds M$_{evol}$=20.0$^{+2.3}_{-1.7}$ \msun\ for the primary of V478~Cyg, in reasonable agreement with \citet{vw10}. The different conclusions reached by us and Vink \& Weidner regarding the discrepancy between evolutionary and dynamical masses is thus not clear. A dedicated analysis of their sample with our method is required to shed more light on this issue.

The right panel of Fig.\ \ref{fig_mass} shows that the uncertainty on the spectroscopic masses are too large to allow any meaningful conclusion regarding the comparison between evolutionary and spectroscopic masses. This is partly due to the difficulty of the disentangling technique to correctly separate the Balmer lines of the systems' components, and thus to the difficulty of determining surface gravities. 

\subsection{Periods and synchronization}
\label{s_T}

Table \ref{tab_perio} summarizes the rotational and orbital periods of each star/system. All systems but DH~Cep being eclipsing, and assuming that the rotational axis is parallel to the orbital axis, the rotational period can be obtained from \vsini\ and the radius obtained from the fit of the light curve. In two systems (Y~Cyg and AH~Cep) the rotational periods are shorter than the orbital one. Since tides tend to synchronize the rotation of stars with the orbital period \citep[e.g.][]{zahn77}, one can conclude that in these two systems the stars have not finished yet to spin-down and that they were initially rotating with equatorial velocities larger then 150 \kms. Interestingly, these systems are also the closest to the ZAMS (together with DH~Cep for which no rotational period can be determined). 

For V382~Cyg both periods are the same within the error bars, indicating that synchronization has been reached. This is also partly true for V478~Cyg. 

XZ~Cep is a peculiar case since the primary has P$_{rotation}$ $<$ P$_{orb}$ while the opposite is true for the secondary. Given that this is the most evolved system of our sample and that the secondary fills its Roche lobe, this could indicate that mass and angular momentum transfer has lead to spin-up of the mass gainer and spin-down of the mass donor. 

A better investigation of the effect of stellar evolution and mass transfer on rotational and orbital periods requires dedicated models. Such a study will be presented in a subsequent paper.

\begin{table}
\begin{center}
\caption{Rotational and orbital periods} \label{tab_perio}
\begin{tabular}{lccc}
\hline
Star         & P$_{rot}$  &  P$_{orb}$  & P$_{rotation}$/P$_{orb}$\\    
             & [d]           &  [d]       &  \\
\hline
V478Cyg-1    & 3.14$\pm$0.33  & 2.88086  & 1.09$\pm$0.11 \\
V478Cyg-2    & 3.51$\pm$0.36  &          & 1.22$\pm$0.13 \\
YCyg-1       & 2.13$\pm$0.19  & 2.99633  & 0.71$\pm$0.06 \\
YCyg-2       & 1.61$\pm$0.14  &          & 0.54$\pm$0.05 \\
V382Cyg-1    & 1.82$\pm$0.12  & 1.88555  & 0.97$\pm$0.06 \\
V382Cyg-2    & 1.83$\pm$0.13  &          & 0.97$\pm$0.07 \\
XZCep-1      & 3.08$\pm$0.19  & 5.09725  & 0.60$\pm$0.02 \\
XZCep-2      & 6.43$\pm$0.66  &          & 1.26$\pm$0.16 \\
AHCep-1      & 1.36$\pm$0.09  & 1.77473  & 0.77$\pm$0.04 \\
AHCep-2      & 1.34$\pm$0.10  &          & 0.76$\pm$0.04 \\
DHCep-1      & $<$2.52       &  2.11095  & $<$1.19\\
DHCep-2      & $<$3.54       &           & $<$1.68 \\
\hline
\end{tabular}
\end{center}
\end{table}

\section{Conclusion}
\label{s_conc}

In this paper we have studied six Galactic massive binaries. We have obtained high-resolution spectroscopy at different orbital phases. A spectral disentangling technique was applied to yield the individual spectra of each component. These spectra were subsequently analyzed with atmosphere models and synthetic spectra to provide the stellar parameters and CNO surface abundances. The results are:

\begin{itemize}

\item[$\bullet$] Five of the six systems have their components located on the main sequence (as defined by single-star evolutionary tracks). For the sixth system, one component is evolved.

\item[$\bullet$] Evolutionary masses determined using the single star evolutionary tracks of \citet{ek12} are on average 16.5\% higher than dynamical masses. 

\item[$\bullet$] Three systems are detached (Y~Cyg, V478~Cyg and AH~Cep), two have at least one star filling its Roche lobe (V382~Cyg and XZ~Cep) and no information is available for DH~Cep since it is not eclipsing.

\item[$\bullet$] The CNO surface abundances of the three detached systems are little affected by stellar evolution and binarity. They are consistent with abundances of single stars within the uncertainties. Similar conclusions apply to the components of the contact system V382~Cyg and to the primary of XZ~Cep. The N/C ratio of the secondary of XZ~Cep is higher than what is expected for single stars with similar initial masses \citep[according to the evolutionary tracks of][]{ek12}.

\item[$\bullet$] Comparison of the derived surface abundances to those of systems known to have experienced mass transfer suggests the following trend: tides do not significantly affect the surface abundances of detached systems, while mass transfer, through the removal of external layers, can make chemically processed material appear at the surface. It may take some time after the onset of mass transfer before such effects are visible.

\item[$\bullet$] Some systems have reached synchronization, while other may still be in a spin-down phase. In the contact binary XZ~Cep mass transfer has probably affected the rotational periods.

\end{itemize}
  
The trends emerging from our results, especially regarding the effect of binarity on surface abundances, need to be confirmed by analysis of additional detached and post-mass transfer systems.

\section*{Acknowledgments}

We thank an anonymous referee for a prompt and positive report. We thank the Observatoire de Haute Provence for assistance with the T-193 telescope and the \emph{SOPHIE} instrument. We warmly thank John Hillier for making the code CMFGEN available to the community and for constant help with it. L.M. acknowledges support from the Fonds National de la Recherche Scientifique (F.R.S.-F.N.R.S.) and through the ARC grant for Concerted Research Actions, financed by the French Community of Belgium (Wallonia-Brussels Federation).

\bibliographystyle{aa}
\bibliography{bin_ohp}

\end{document}